\documentclass[a4paper]{cas-sc}

\usepackage[numbers]{natbib}
\usepackage{xcolor}
\usepackage{booktabs}
\usepackage{colortbl}
\usepackage{subcaption}

\usepackage{graphicx}
\usepackage{float} 
\usepackage{placeins}
\usepackage{url}
\usepackage{hyperref}

\usepackage{rotating}
\usepackage{array}

\definecolor{risk}{RGB}{250,150,120}
\definecolor{warning}{RGB}{248,223,146}
\definecolor{safe}{RGB}{178,220,166}
\definecolor{headerblue}{RGB}{70,75,120}

\def\tsc#1{\csdef{#1}{\textsc{\lowercase{#1}}\xspace}}
\tsc{WGM}
\tsc{QE}
\tsc{EP}
\tsc{PMS}
\tsc{BEC}
\tsc{DE}
\begin{document}
\let\WriteBookmarks\relax
\def\floatpagepagefraction{1}
\def\textpagefraction{.001}

\shorttitle{HPAI Modelling on Jolly Island}

\shortauthors{Fatoyinbo et~al.}

\title [mode = title]{Stochastic Spatial Metapopulation Modelling of HPAI Control and Poultry Restocking on Jolly Island}                      
%\tnotemark[1,2]

%\tnotetext[1]{This document is the results of the research
%   project funded by the National Science Foundation.}

%\tnotetext[2]{The second title footnote which is a longer text matter
 %  to fill through the whole text width and overflow into
  % another line in the footnotes area of the first page.}

\author[1]{Hammed O. Fatoyinbo}[orcid=0000-0002-6036-2957]

\cormark[1]


\ead{hammed.fatoyinbo@aut.ac.nz}

%\ead[url]{www.cvr.cc, cvr@sayahna.org}

\credit{Writing – review \& editing, Writing – original draft, Visualization, Validation, Supervision, Software, Resources, Project administration, Methodology, Investigation, Formal analysis, Data curation, Conceptualization}

\affiliation[1]{organization={Department of Mathematical Sciences, Auckland University of Technology},
    %addressline={Radarweg 29}, 
    city={Auckland},
    % citysep={}, % Uncomment if no comma needed between city and postcode
    postcode={1010}, 
    country={New Zealand}}

\author[2]{Indranil Ghosh}[orcid=0000-0001-8078-0577]
\ead{indra.ghosh@ucd.ie}
\credit{Writing – review \& editing, Writing – original draft, Visualization, Validation, Software, Resources, Project administration, Methodology, Investigation, Formal analysis, Data curation, Conceptualization}

\affiliation[2]{organization={School of Mathematics and Statistics, University College Dublin},
    %addressline={Radarweg 29}, 
    city={Dublin},
    % citysep={}, % Uncomment if no comma needed between city and postcode
    postcode={4-D04 V1W8}, 
    country={Ireland}}

\author[1]{Parul Tiwari}[orcid=0000-0003-0206-6604
   ]
\ead{parul.tiwari@aut.ac.nz}
%\ead[URL]{www.sayahna.org}

\credit{Writing – review \& editing, Writing – original draft, Visualization, Validation, Software, Resources, Methodology, Investigation, Formal analysis, Data curation, Conceptualization}

%\affiliation[1]{organization={Department of Mathematical Sciences, Auckland University of Technology},
    %addressline={Radarweg 29}, 
  %  city={Auckland},
    % citysep={}, % Uncomment if no comma needed between city and postcode
  %  postcode={1010}, 
   % country={New Zealand}}

\author%
[3]
{Peter O. Olanipekun}[orcid=0000-0002-3310-1655]
%\cormark[2]
%\fnmark[3]
\ead{olanipekunp@gmail.com}
%\ead[URL]{www.stmdocs.in}

\credit{Writing – review \& editing, Writing – original draft, Visualization, Validation, Software, Resources, Methodology, Investigation, Formal analysis, Data curation, Conceptualization}

\affiliation[3]{organization={Department of Mathematics, University of Auckland},
    %addressline={Radarweg 29}, 
    city={Auckland},
    % citysep={}, % Uncomment if no comma needed between city and postcode
    postcode={1010}, 
    country={New Zealand}}

% Fifth author
\author%
[4]
{Afeez Abidemi}[orcid=0000-0003-1960-0658]
%\cormark[2]
%\fnmark[3]
\ead{aabidemi@futa.edu.ng}
%\ead[URL]{www.stmdocs.in}
\credit{Writing – review \& editing, Writing – original draft, Validation, Methodology, Investigation, Formal analysis, Data curation, Conceptualization}

\affiliation[4]{organization={Department of Mathematical Sciences, Federal University of Technology Akure},
    %addressline={}, 
   % city={Akure},
    % citysep={}, % Uncomment if no comma needed between city and postcode
    %postcode={}, 
    state={Ondo State},
    country={Nigeria}}

\author%
[1]
{Ryan H.L. Ip}[orcid=0000-0001-8636-1891]
%\fnmark[3]
\ead{ryan.ip@aut.ac.nz}
%\ead[URL]{www.stmdocs.in}
\credit{Writing – review \& editing, Writing – original draft, Visualization, Validation, Software, Resources, Methodology, Investigation, Formal analysis, Data curation, Conceptualization}
%\affiliation[1]{organization={Department of Mathematical Sciences, Auckland University of Technology},
%%   country={New Zealand}}

\cortext[cor1]{Corresponding author}
%\cortext[cor2]{Principal corresponding author}

%\fntext[fn1]{This is the first author footnote. but is common to third
 % author as well.}
%\fntext[fn2]{Another author footnote, this is a very long footnote and
 % it should be a really long footnote. But this footnote is not yet
 % sufficiently long enough to make two lines of footnote text.}

%\nonumnote{This note has no numbers. In this work we demonstrate $a_b$
%  the formation Y\_1 of a new type of polariton on the interface
 % between a cuprous oxide slab and a polystyrene micro-sphere placed
 % on the slab.
 % }

\begin{abstract}
Highly pathogenic avian influenza (HPAI) outbreaks require rapid control during active transmission and evidence-based decisions on the safe restocking of depopulated farms. We developed a stochastic spatial SEIR-based metapopulation model for a synthetic HPAI outbreak on the fictional Jolly Island. Farms were classified as `Broiler-2', `organic duck', or `Other' production systems. The model incorporated local, environmental, movement-mediated, and distance-dependent transmission, together with reactive and preventive culling, production-specific confinement, and capacity-based restocking. The simulated epidemic was geographically concentrated and differed substantially among production classes. Preventive culling reduced mean cumulative burden from 16,362.7 to 13,631.9 infectious-farm-days, with an overall reduction of 16.7\%. Earlier confinement substantially reduced epidemic magnitude, while stronger environmental transmission increased the epidemic peak. Restocking risk declined as the epidemic approached resolution. Under the model assumptions, 24 May 2026 was the first candidate date satisfying the predefined rebound-probability threshold of 0.20. For restocking on 15 March 2026, none of the tested restocking fractions met this criterion. Capacity-based restocking reduced cumulative burden by 8.45\% and rebound probability from 0.780 to 0.533, compared with restocking relative to the baseline population. These findings demonstrate the value of integrating epidemic control and post-outbreak recovery within a single modelling framework. Timely confinement, targeted preventive culling, and phased capacity-based restocking may reduce both epidemic burden and resurgence risk, although operational decisions should also incorporate surveillance, biosecurity, economic considerations, and regulatory requirements.
\end{abstract}

% \includegraphics{figs/grabs.pdf}

%\item Research highlights item 1
%\item Research highlights item 2
%\item Research highlights item 3

% Keywords

\maketitle

\section{Introduction}
Highly pathogenic avian influenza (HPAI) has re-emerged as a major transboundary threat to poultry production, animal health, food security and public health. The recent expansion of clade 2.3.4.4b H5Nx viruses has intensified concern because these viruses circulate widely in wild birds, repeatedly spill over into domestic poultry, and have increasingly been detected in mammals, including carnivores and dairy cattle \citep{BRIAND2026105958,Charostad2023,WilleBarr2022,Leguia2023,Vreman2023,Butt2024,Burrough2024,Caserta2024,OwusuSanad2025}. Although the principal burden of HPAI remains concentrated in avian populations, the widening host range and repeated cross-species incursions have reinforced the need for integrated One-Health surveillance, rapid outbreak response and quantitative tools capable of evaluating control strategies under uncertainty \citep{LancetInfectiousDiseases2024,Duarte2024,Sah2024,Rawson2025}. 

During an outbreak, rapid and effective responses are especially important for poultry-producing regions since the epidemiological and economic consequences can be severe: infected premises are commonly depopulated, farms in high-risk neighbourhoods may be subjected to preventive culling, movements may be restricted, and restocking may be delayed until the risk of recrudescence is sufficiently low \citep{WOAH2024,CDC2025,USDA2025}.

In addition, the transmission ecology of HPAI is inherently multi-scale. At broad spatial scales, wild-bird migration and environmental interfaces shape the risk of introduction into poultry farms \citep{Olsen2006,Kilpatrick2006,Caliendo2022,Stanislawek2024,Esaki2025, Fato2025, HamHoy26}. At regional and local scales, spread among poultry farms depends on farm density, production type, indirect contacts, movements of birds and materials, service vehicles, personnel, equipment sharing and biosecurity practices \citep{Tiensin2007,Fasina2011,Belkhiria2016,Dunning2025}. Environmental persistence further complicates control, because avian influenza viruses can remain infectious in water or contaminated substrates for durations that depend on temperature, salinity, pH and other environmental conditions \citep{Stallknecht1990,Brown2009,Lebarbenchon2010}. Consequently, the force of infection experienced by a susceptible poultry farm may arise from multiple concurrent pathways such as local neighbourhood transmission, movement-mediated contacts, environmental exposure and long-distance transmission. Models intended for outbreak decision support must therefore take into account more than a single homogeneous mixing process.

Mathematical modelling has long played a central role in linking epidemic dynamics to the design and evaluation of control policies. Classical compartmental models clarify how susceptible, exposed, infectious and removed populations interact, while stochastic formulations are particularly important for livestock epidemics in which early transmission events, local extinction and between-farm heterogeneity can strongly influence outcomes \citep{AndersonMay1991,KeelingRohani2008,Allen2008,GarnerHamilton2011}. Spatial, network and metapopulation models extend this framework by allowing infection pressure in one location to depend on disease states elsewhere through distance kernels, movement networks or other structured contact processes \citep{Riley2007,Colizza2007,Balcan2009}. Such approaches have been used extensively in animal-disease epidemiology, including foot-and-mouth disease and avian influenza, to quantify the effect of farm density, spatial clustering, delayed detection and reactive or preventive culling \citep{Keeling2001,Ferguson2001,Tildesley2010,Boender2007,Dorigatti2010,Sharkey2008}. For HPAI, this modelling tradition is especially relevant because epidemic outcomes are strongly influenced by where infected farms are located, how farms are connected, and how quickly infectious premises are detected and removed.

Recent avian-influenza modelling studies have increasingly emphasised explicit spatial structure, multiple host or production groups, environmental reservoirs and intervention evaluation. Mechanistic models have been used to investigate avian-human transmission, spread between migratory birds and domestic poultry, fractional-order avian-human systems, and optimal control formulations for delayed HPAI dynamics \citep{Iwami2007,Liu2008,Bourouiba2011,Sharma2018,Ye2020}. Data-driven spatial analyses have examined HPAI transmission risk in wild birds and poultry, including risk mapping, multi-criteria geospatial surveillance and regional transmission patterns \citep{Pandit2013,Sangrat2024,Fatoyinbo2025}. A recent systematic review concluded that avian-influenza models vary substantially in epidemiological unit, spatial scale, transmission route and control representation, and highlighted the need for models that can evaluate practical interventions in real time \citep{Lambert2023}. Recent studies have also address HPAI risk outside poultry, including spillover into cattle and multi-host environmental transmission, underscoring the importance of flexible modelling frameworks that can represent transmission pathways beyond direct bird-to-bird contact \citep{Hassman2025,Chang2026,Rawson2025}.

Despite this progress, several operational gaps remain. First, many outbreak-response questions must be answered using partially observed, aggregated and rapidly changing data rather than complete farm-level transmission histories. Second, decision-makers often need to compare interventions that act on different mechanisms. For instance, reactive culling shortens the infectious period of detected farms, preventive culling removes farms that may be at high risk, confinement alters selected contact or exposure routes, and movement restrictions modify network-mediated transmission. Third, epidemic management does not end when incidence declines. Depopulated farms must eventually be restocked, but premature reintroduction of susceptible poultry can reignite transmission if infectious farms or contaminated environments remain. Restocking decisions therefore require probabilistic risk metrics, not only deterministic forecasts of incidence. These issues motivate stochastic, spatially structured decision-support models that can integrate observed outbreaks, production systems, movement data, culling schedules, environmental persistence and post-control recovery.

In this paper, we develop a stochastic spatial SEIR-based metapopulation model for a synthetic HPAI outbreak on Jolly Island as part of the HPAI Modelling Challenge organised by WiLiMan-ID \citep{wiliman26}. The full challenge dataset describes an epidemic that happened on a poultry-producing island with 14 counties, 556 districts, five poultry production systems, a designated high-risk zone along part of the eastern coast, recorded farm movements, confirmed outbreaks, reactive culling, preventive culling and confinement interventions. Across the complete observation period between 20 December 2025 and 7 April 2026, 560 confirmed outbreaks were recorded, with outbreaks concentrated in chicken farms and especially broiler production systems. The epidemic initially appeared along the eastern coast, developed northern and southern clusters, expanded westward, and eventually declined, with residual cases concentrated mainly in northern areas. These features indicate a spatially heterogeneous epidemic shaped by production structure, local clustering and inter-county spread.

The model considers farms as epidemiological units aggregated by county and production class. Within each county-production stratum, farms are classified as susceptible, exposed, infectious or removed, and each county has a shared environmental contamination compartment. The total force of infection is decomposed into four components: local within-county transmission, environmental transmission, movement-mediated transmission and spatial transmission through an exponential distance-decay kernel. This decomposition follows the biological and operational structure of HPAI transmission in the sense that infected farms can transmit locally, shed virus into the environment, export risk through recorded movements and contribute to spatial infection pressure in nearby counties. An important feature of the proposed model is that control measures are embedded directly in the stochastic process. Reactive and preventive culling remove farms according to the recorded intervention schedules, confinement modifies selected transmission pathways, and restocking reintroduces susceptible farms into depopulated capacity at candidate dates.

The main contribution of this paper is a decision-oriented stochastic metapopulation framework for evaluating HPAI control and recovery under uncertainty. We use the model to reconstruct the epidemic trajectory, quantify the effect of preventive culling, and assess when depopulated farms could be restocked without causing a rebound. Preventive culling is evaluated by comparing cumulative infectious burden under scenarios with and without recorded preventive culls. Restocking is evaluated through weekly candidate dates, with each date classified according to the simulated probability that post-restocking infections exceed the infection level at restocking by more than a specified tolerance. This produces directly interpretable quantities (such as expected infectious burden, outbreaks averted, rebound probability and safe restocking windows) for outbreak management.

The remainder of the paper is organised as follows. Section 2 describes the challenge data, epidemic summary, county-production SEIR model, environmental contamination process, movement and spatial transmission terms, culling implementation and restocking-risk formulation. Section 3 reports the parameter values and scenario outcomes. Section 4 interprets the implications of the model for HPAI control, preventive culling and restocking, and outlines the limitations associated with county-level aggregation, homogeneous farms within strata and fixed parameter values. Finally, Section 5 summarises the value of stochastic spatial metapopulation models as rapid decision-support tools for HPAI outbreak management.

\section{Materials and methods}
\label{sec:materials}

\subsection{Challenge Data and Control Strategies}
\begin{figure}[pos=htbp]
    \centering
    \includegraphics[width=.4\columnwidth]{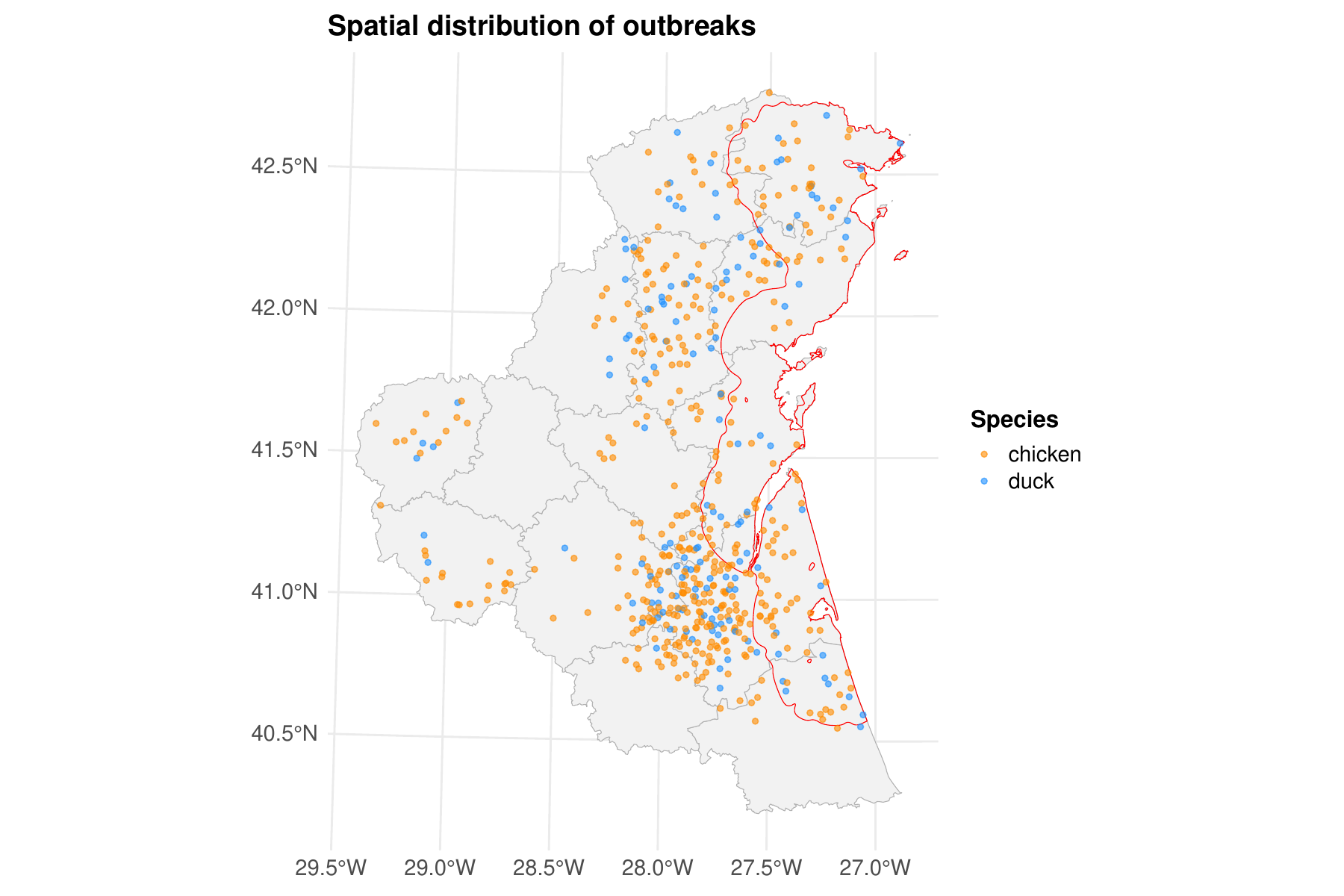}
    \includegraphics[width=.4\columnwidth]{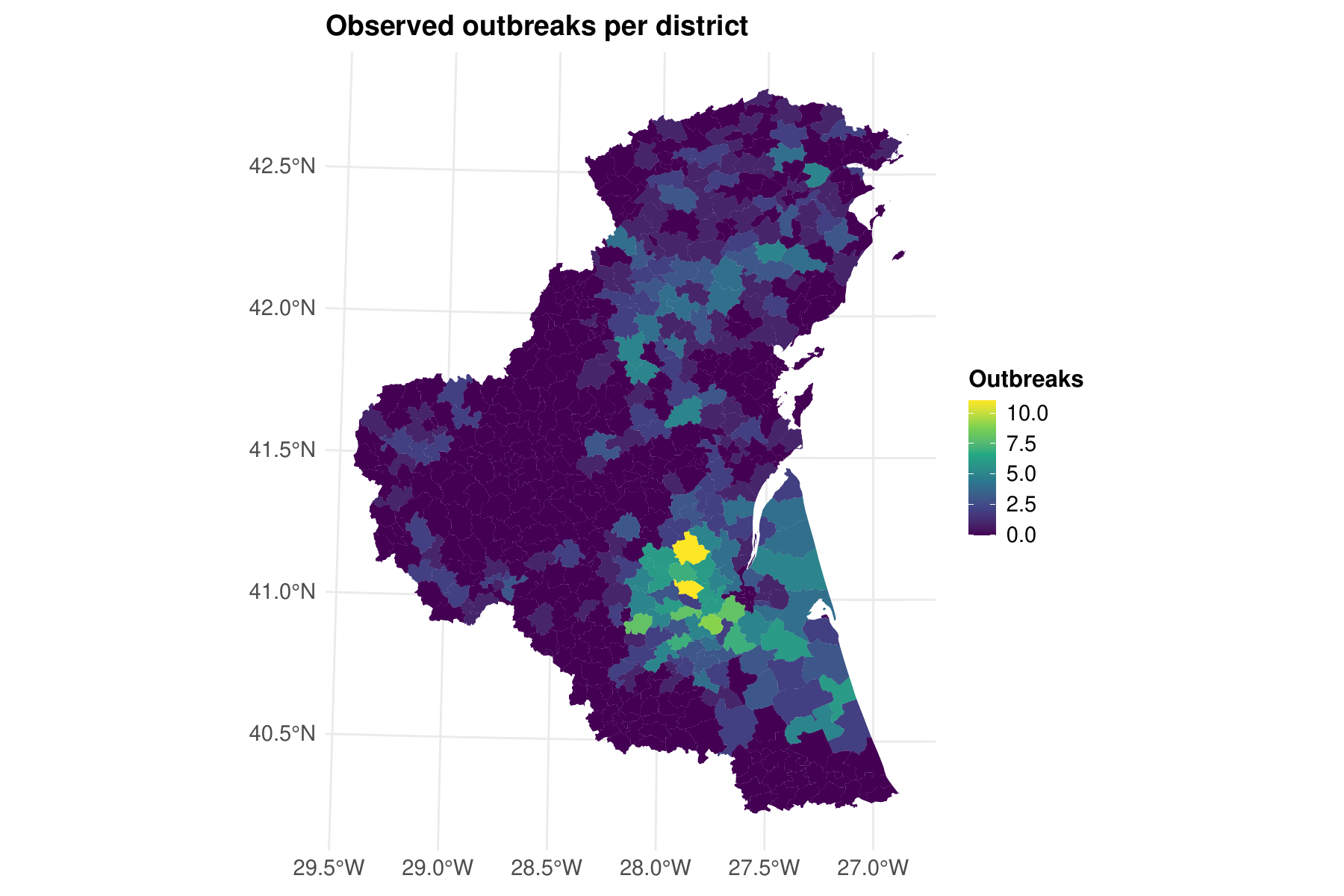}
    \caption{Maps of the study region (Jolly Island). The left panel shows the counties (grey boundaries), the high-risk zone (the red boundary), and the locations of HPAI outbreaks by species over the study period. The right panel shows the district-level outbreak frequencies.}
    \label{fig_maps}
\end{figure}

In this challenge, a set of synthetic epidemic data was provided by the challenge coordinators, which concerned a new HPAI viral strain detected on Jolly Island, a made-up country located in the Atlantic Ocean. Jolly Island consisted of 14 counties and 556 districts, and covered an area of over 33,000 km$^2$. A map of the study region is provided in Figure \ref{fig_maps}. The main farm animal production was poultry, with laying hens, Stage 1 and 2 broiler chickens, organic meat ducks and conventional meat ducks being the main production systems. Due to the movement of migratory birds, part of the east coast of the island was categorised as the high-risk zone. 

The first suspicion was reported on 20 December 2025, and was later confirmed on 22 Dec 2025. By the end of the year (31 Dec 2025), 8 outbreaks were confirmed. The data were released in three phases, mimicking the real-life situation where data arrive in batches during an epidemic. Phase 1 contained the data from 20 Dec 2025 to 13 Jan 2026. Phase 2 data extended the timeline to 13 Feb 2026, and Phase 3 data covered all cases until 7 Apr 2026. In other words, the full dataset covers a total of 109 days, considering 20 Dec 2025 as Day 1.

Apart from farms where an outbreak was detected, data about between-farm movements and culling, a controlling strategy, were also available. Reactive culling was performed in confirmed farms. Besides, preventive culling was initiated on 1 Jan 2026, after the initial surveillance strategy failed to control the epidemic. In the first phase, all animals in poultry farms located within 1 km of the confirmed farms were culled. In phases 2 and 3, the radius extended to 3 km. The data provided detailed information of the dates when culling was planned and completed. Confinement was another major control strategy introduced. In Phase 2, confinement was introduced to stage 2 broiler and organic duck farms, although the regulation was later lifted for duck farms in Phase 3.

\subsection{Summary of the Epidemic}
\begin{figure}[pos=htbp]
	\centering
		\includegraphics[width=.55\columnwidth]{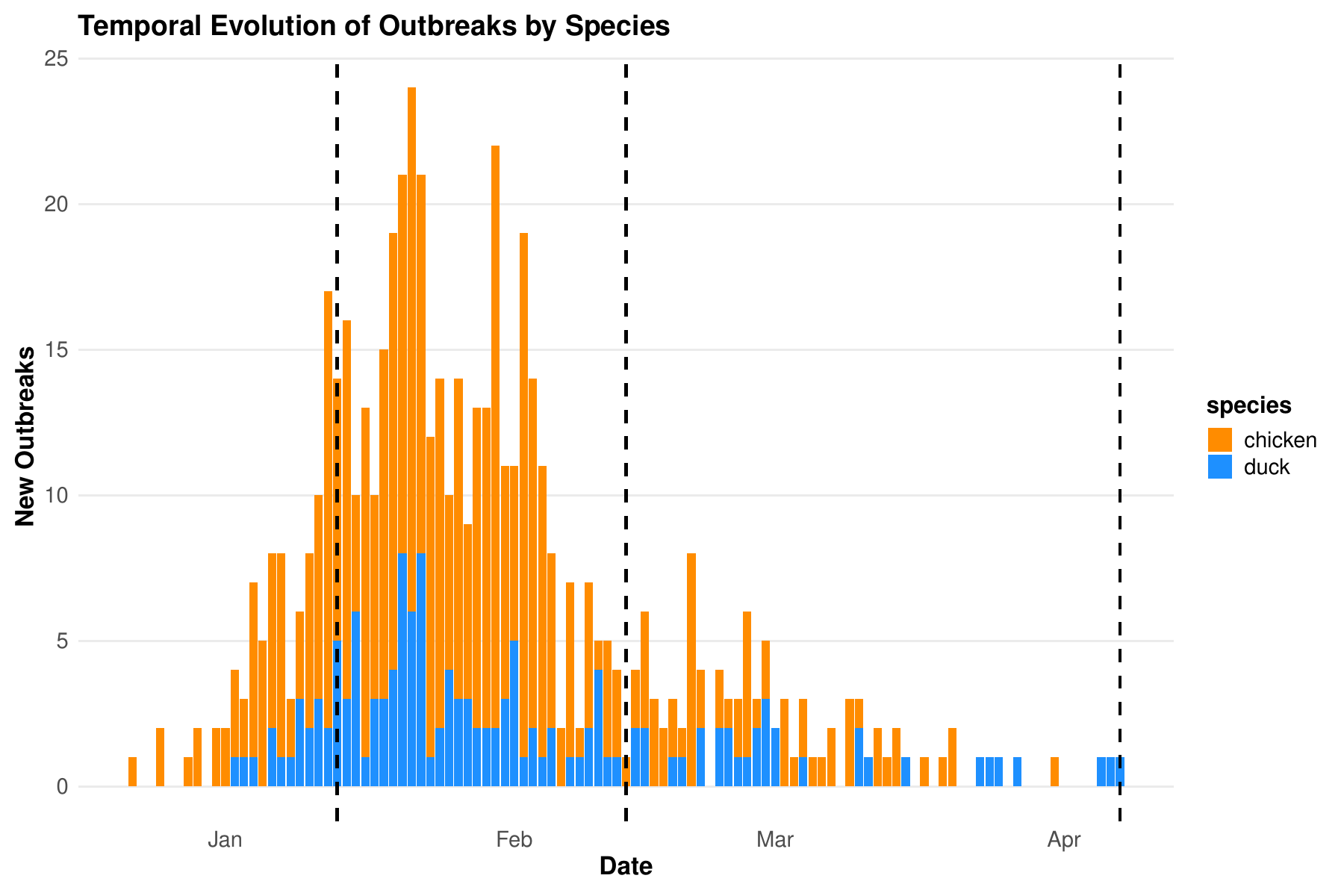}
	\caption{Number of daily new confirmed HPAI cases by species over the study period. The three vertical dashed lines indicate the end of the three data release phases. }
	\label{fig_temporal_species}
\end{figure}

Over the entire period, 560 outbreaks were observed, with 420 (75\%) occurring in chicken farms. From Figure \ref{fig_temporal_species}, it can be seen that most of the outbreaks were observed between mid-January and mid-February, with the peak occurring on 21 Jan 2026 when 24 new cases were confirmed. 

\begin{figure}[pos=htbp]
	\centering
		\includegraphics[width=.65\columnwidth]{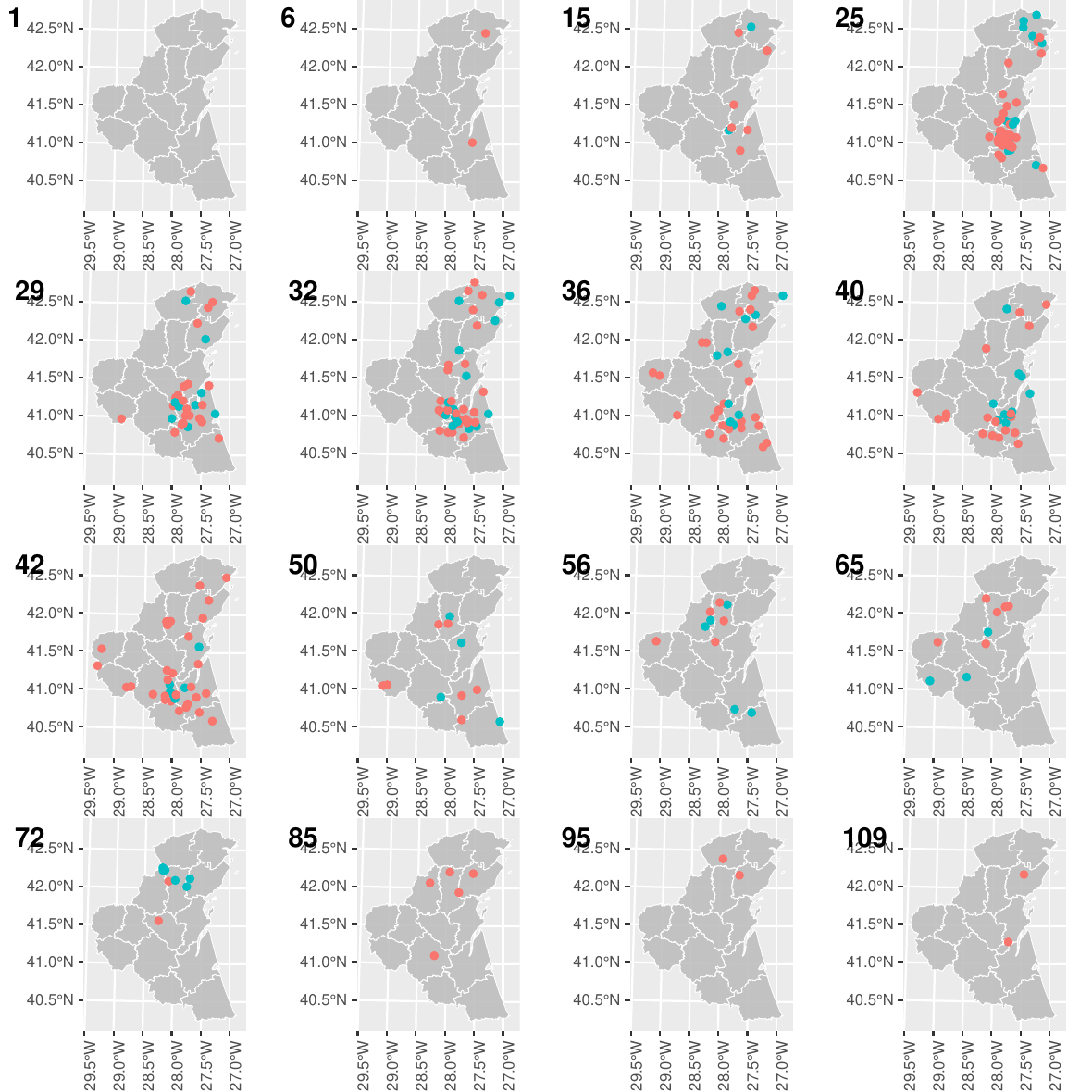}
	\caption{Locations of infected farms across a selection of days over the study period. The days since the first suspicion (20 Dec 2025) are labelled at the top-left of each plot. Chicken farms are labelled in orange while duck farms are labelled in green.}
	\label{fig_spatial_temporal}
\end{figure}

Figure \ref{fig_spatial_temporal} shows the location of infected farms on some selected dates between the date the first suspicious case was
reported (20 Dec 2025; Day 1) and the end of the data collection period (07 Apr 2026; Day 109). As seen from the figure, the first confirmed cases were found on the East coast, which coincided with the high-risk zone. Although reactive culling was completed on Day 5, the virus still spread to two other farms, one in the North and one in the South, on Day
6. Since then, the virus eventually spread across the Eastern half of the island within the first 25 days, which mainly consisted of two clusters, one in the North and one in the South.

   The first confirmed case along the West coast was
reported on Day 29, and more cases appeared after Day 36 in that area. Meanwhile, the situation in the
Central and North areas improved, with the number of confirmed cases continuing to decline until the end of the second data release phase (Day 56). Since the beginning of the third phase, the number of cases continued to decline in all areas. The majority of the remaining cases were located in the northern areas. Towards the end of the data collection period, only a few infected farms remained, scattered on the Island. Throughout the entire study period, most of the infected farms were breeding chickens (orange), compared to ducks (green).

\subsection{Modelling framework and analyses}\label{sec:modelframework}

We developed a stochastic, spatially explicit, discrete-time SEIR metapopulation model to simulate the transmission of HPAI among poultry farms on Jolly Island. The model was structured at the county level, with counties represented as epidemiological patches connected through recorded farm movements and spatial proximity. Within each county, farms were stratified into three production classes: stage 2 broiler chicken farms, hereafter referred to as Broiler-2 farms, organic duck farms, and all other farms, denoted by $B2$, $D$, and $O$, respectively.

The model represented farms within each county--production stratum as susceptible, exposed, infectious, or removed. It also included a county-level environmental contamination compartment to represent indirect transmission through contaminated surroundings. Transmission was represented through four pathways: local within-county transmission, environmental exposure, movement-mediated transmission based on recorded farm movements, and distance-dependent spatial transmission between counties. The model further represented reactive culling, preventive culling, confinement, and restocking.

Because the model was formulated at the county--production level, farms within the same county and production class were assumed to be epidemiologically homogeneous. Farms within a stratum therefore shared the same transmission parameters, susceptibility, and disease-progression probabilities. Individual farm-level variation in flock size, biosecurity, management practices, contact behaviour, and susceptibility was not explicitly represented. Environmental contamination was shared across production classes within each county.

Movement-mediated transmission was based on recorded farm movements aggregated to the county level. Unrecorded movements and other indirect contacts, including movements of personnel, equipment sharing, service vehicles, and informal contacts, were not explicitly modelled. Their aggregate effects may nevertheless have been partially represented by the local, environmental, or spatial transmission components.

The epidemic was seeded once at the beginning of the simulation using the first five confirmed farms in chronological order. No additional background seeding was included in the main analyses. Simulated epidemic dynamics were therefore generated by the initial infections and subsequent transmission processes rather than repeated external introductions. 

Culling was implemented as a scheduled removal process using recorded reactive and preventive culling dates. Culling was applied after the daily transmission and disease-progression transitions. Because the disease state of each scheduled culled farm was not tracked individually in the aggregated model, the required number of farms was subtracted sequentially from the susceptible, exposed, infectious, and removed compartments. This sequence was an accounting rule for applying recorded culling totals and should not be interpreted as an epidemiological prioritisation of susceptible farms.

Restocked farms were assumed to be susceptible and uninfected at reintroduction. The primary restocking analysis used a capacity-based formulation in which the number of farms reintroduced was constrained by the empty capacity created by previous removals. This ensured that the total number of farms in each county--production stratum could not exceed its baseline population.

The main components of the modelling framework were as follows:

\begin{itemize}

    \item[i.] \textbf{Production-type classification:}
    Farms were grouped into three production classes, $p$. Chicken farms recorded as Broiler-2 were assigned to the $B2$ class, duck farms recorded as organic were assigned to the $D$ class, and all remaining farms were assigned to the $O$ class. Thus,
     \begin{equation*}
    p \in \{B2,D,O\}.
    \end{equation*}
  For each county $c$ and production class $p$, farms were aggregated to define the baseline county--production population,
    \[
    N_{c,p}.
    \]
   % where $N_{c,p}$ denotes the number of farms in county $c$ and production class $p$.

    \item[ii.] \textbf{State variables:}
    For each county $c$, production class $p$, and day $t$, farms were assigned to one of four epidemiological compartments:
    \[
    S_{c,p}(t), \quad E_{c,p}(t), \quad I_{c,p}(t), \quad R_{c,p}(t),
    \]
   where $S_{c,p}(t)$ denotes susceptible farms, $E_{c,p}(t)$ denotes exposed farms, $I_{c,p}(t)$ denotes infectious farms, and $R_{c,p}(t)$ denotes farms that had completed the infectious process and no longer contributed to transmission. Scheduled culling was subsequently applied as an additional subtraction from these county--production state variables.

    The model also included a county-level environmental contamination compartment,
    \[
    Env_c(t),
    \]
   representing the accumulated environmental contamination in county $c$. This environmental compartment was shared across production classes within each county.

    \item[iii.] \textbf{Simulation period and initial conditions:}
  The simulation ran from 22 December 2025 to 30 June 2026 using a daily time step. Observed outbreak and intervention records were available through 7 April 2026 while the simulation was extended to 30 June 2026 to evaluate epidemic resolution and potential restocking dates beyond the observation period. 
  
  At the beginning of the simulation, all farms were assumed susceptible except for the initial seeded infections. The first five confirmed farms in chronological order were introduced as infectious at $t=0$. For each seeded farm, one farm was transferred from the susceptible compartment to the infectious compartment in the corresponding county--production stratum:
    \[
    S_{c,p}(0) \leftarrow S_{c,p}(0)-1,
    \]
    \[
    I_{c,p}(0) \leftarrow I_{c,p}(0)+1.
    \]
    No additional background seeding was included in the main analyses.

    \item[iv.] \textbf{High-risk zone adjustment:}
    County-level hazard multipliers were calculated as the mean farm-level multiplier within each county:
    \begin{equation*}
        h_c =
    \frac{1}{N_c}
    \sum_{i \in c} h_i,
    \end{equation*}
    
    where $h_i$ is the hazard multiplier assigned to farm $i$, $N_c$ is the total number of farms in county $c$, and $h_c$ is the resulting county-level multiplier. The multiplier $h_c$ was applied to the environmental transmission component.

    Farms located inside the designated high-risk zone were assigned a multiplier of 1.5, whereas farms outside the zone were assigned a multiplier of 1.0. If overlapping polygons occurred, a farm was classified as being inside the high-risk zone if it was contained within at least one high-risk zone polygon. 

    \item[v.] \textbf{Movement network:}
    Movement-mediated transmission was represented using the daily movement tensor \[ M_{t,c',c,p}, \] where $M_{t,c',c,p}$ denotes the recorded movement volume on day $t$ from source county $c'$ to destination county $c$ for production class $p$. Movements were assigned to a production class according to the classification of the source farm.

    The infectious prevalence in source county $c'$ and production class $p$ is defined as
    \[
    \pi_{c',p}(t)
    =
    \frac{I_{c',p}(t)}
    {N_{c',p}+\varepsilon},
    \]
    where $\varepsilon$ is a small constant included to avoid division by zero. In this work, we have set $\varepsilon=10^{-6}$.

    \item[vi.] \textbf{Spatial transmission kernel:}
    Spatial transmission between counties was modelled using an exponential distance-decay kernel based on county centroid distances. Let $d_{c,c'}$ denote the Euclidean distance between the centroids of counties $c$ and $c'$. The spatial kernel is defined as
    \[
    K_{c,c'}
    =
    \exp
    \left(
    -\frac{d_{c,c'}}{d_0}
    \right).
    \]
    
    where the spatial decay parameter $d_0$ was set to $2000\ \mathrm{m}$. No hard distance cut-off was imposed. Consequently, $K_{c,c'}$ remained positive when $d_{c,c'}>d_0$, although its value declined exponentially with increasing distance. The diagonal elements were set to zero (i.e. $K_{c,c}=0$), to exclude self-transmission through the between-county spatial kernel.

   \item[vii.] \textbf{Transmission pressure:} The total transmission pressure for county $c$, production class $p$, and day $t$ was defined as the sum of four components: 
   \[ \lambda_{c,p}(t) = \lambda^{\mathrm{local}}_{c,p}(t) + \lambda^{\mathrm{env}}_{c,p}(t) + \lambda^{\mathrm{move}}_{c,p}(t) + \lambda^{\mathrm{spatial}}_{c,p}(t). \] 
   
   Local transmission within the same county and production class is given as
   \[ \lambda^{\mathrm{local}}_{c,p}(t) = \beta^{\mathrm{within}}_p I_{c,p}(t), \]
   where $\beta^{\mathrm{within}}_p$ is the production-specific local transmission coefficient. 
   
   Environmental transmission pressure is given as
   \[ \lambda^{\mathrm{env}}_{c,p}(t) = \beta^{\mathrm{env}}_p Env_c(t) h_c m^{\mathrm{exp}}_p(t), \]
   where $\beta^{\mathrm{env}}_p$ is the environmental transmission coefficient and $m^{\mathrm{exp}}_p(t)$ is the confinement multiplier applied to environmental exposure. The confinement multipliers will be explained below. 
   
   Movement-mediated transmission pressure is given as
   \[ \lambda^{\mathrm{move}}_{c,p}(t) = \beta^{\mathrm{move}}_p m^{\mathrm{mov}}_p(t) \sum_{c'} M_{t,c',c,p} \pi_{c',p}(t), \]
   where $\beta^{\mathrm{move}}_p$ is the movement-mediated transmission coefficient and $m^{\mathrm{mov}}_p(t)$ is the confinement multiplier applied to movement-mediated transmission. 
   
   Spatial transmission pressure is given as
   \[ \lambda^{\mathrm{spatial}}_{c,p}(t) = \beta^{\mathrm{spatial}}_p \sum_{c'} K_{c,c'} \sum_{p'} I_{c',p'}(t), \]
   where $\beta^{\mathrm{spatial}}_p$ is the production-specific spatial transmission coefficient and $p'$ indexes the production class in the source county. Infectious farms from all production classes in neighbouring counties contributed to the incoming spatial pressure.
   
    \item[viii.] \textbf{Stochastic infection process:}
    The number of newly exposed farms was generated using a Poisson process:
    \[
    newE_{c,p}(t)
    \sim
    \mathrm{Poisson}
    \left(
    \lambda_{c,p}(t)
    \frac{S_{c,p}(t)}
    {N_{c,p}+\varepsilon}
    \right).
    \]
    The number of new exposed farms was truncated so that it could not exceed the number of susceptible farms:
    \[
    newE_{c,p}(t)
    \leq
    S_{c,p}(t).
    \]

    \item[ix.] \textbf{Disease progression and removal:}
    Progression from exposed to infectious status was modelled as
    \[ newI_{c,p}(t) \sim \mathrm{Binomial} \left( E_{c,p}(t),\sigma \right), \]
    and removal from the infectious compartment was modelled as
    \[ newR_{c,p}(t) \sim \mathrm{Binomial} \left( I_{c,p}(t),\gamma \right), \]
    where $\sigma$ is the daily exposed-to-infectious transition probability and $\gamma$ is the daily infectious-to-removed transition probability.
    
    Before scheduled culling, the epidemiological compartments were updated as \[ S_{c,p}(t+1) = S_{c,p}(t) - newE_{c,p}(t), \] 
    \[ E_{c,p}(t+1) = E_{c,p}(t) + newE_{c,p}(t) - newI_{c,p}(t), \] 
    \[ I_{c,p}(t+1) = I_{c,p}(t) + newI_{c,p}(t) - newR_{c,p}(t), \] 
    \[ R_{c,p}(t+1) = R_{c,p}(t) + newR_{c,p}(t). \]

    \item[x.] \textbf{Environmental contamination dynamics:}
    Environmental contamination was updated daily as
    \[
    Env_c(t+1)
    =
    (1-\delta_{env})Env_c(t)
    +
    \sum_p
    \kappa^{env}_p
    m^{shed}_p(t)
    I_{c,p}(t),
    \]
    where $\delta_{env}$ is the daily environmental decay rate, $\kappa^{env}_p$ is the production-specific environmental shedding coefficient, and $m^{shed}_p(t)$ is the confinement multiplier affecting environmental shedding. 

    \item[xi.] \textbf{Reactive and preventive culling:}
    Reactive culling was derived from the confirmed case dataset. Farms with completed culling records were mapped to their county and production class. Preventive culling was derived from the preventive culling dataset. Four preventive culling policies were allowed: no preventive culling, chicken-only preventive culling, duck-only preventive culling, and all recorded preventive culling. The baseline preventive culling scenario used is the all recorded preventive culling.

    At each time step, reactive and preventive culls were combined:
    \[
    C_{c,p}(t)
    =
    C^{reactive}_{c,p}(t)
    +
    C^{preventive}_{c,p}(t).
    \]
    Reactive culling dates could be shifted by a specified delay in sensitivity analyses, whereas preventive culling was applied on its recorded completion date.
    
    Culling was applied after infection and disease progression updates. Farms were removed from the compartments in the order of $S$, $E$, $I$, and $R$ until the scheduled number of culls for the county-production group had been reached.

    \item[xii.] \textbf{Confinement:}
    Confinement was represented using production-specific multipliers applied to environmental exposure, environmental shedding, and movement-mediated transmission. Before the confinement start date ($t_{conf}$), all multipliers were set to one. After confinement was implemented, they were defined as
    
    \[ m^{\mathrm{exp}}_p(t) = \begin{cases} 1, & t<t_{\mathrm{conf}},\\ \widetilde{m}^{\mathrm{exp}}_p, & t\geq t_{\mathrm{conf}}, \end{cases} \]
    
    \[ m^{\mathrm{shed}}_p(t) = \begin{cases} 1, & t<t_{\mathrm{conf}},\\ \widetilde{m}^{\mathrm{shed}}_p, & t\geq t_{\mathrm{conf}}, \end{cases} \]
    
    \[ m^{\mathrm{mov}}_p(t) = \begin{cases} 1, & t<t_{\mathrm{conf}},\\ \widetilde{m}^{\mathrm{mov}}_p, & t\geq t_{\mathrm{conf}}. \end{cases} \] 
    
    In the baseline configuration, the post-confinement environmental exposure multipliers were
    \[ \widetilde{m}^{\mathrm{exp}}_p=(0,0,1), \] 
    
    in the order $(B2,D,O)$. Thus, confinement completely interrupted the modelled environmental exposure pathway for Broiler-2 and organic duck farms, but did not modify environmental exposure for other farms. Environmental shedding and movement-mediated transmission were not modified in the baseline configuration: 
    \[ \widetilde{m}^{\mathrm{shed}}_p=(1,1,1), \qquad \widetilde{m}^{\mathrm{mov}}_p=(1,1,1). \] 
    
    Alternative confinement dates, environmental transmission strengths, and confinement configurations were evaluated in sensitivity analyses.

   \item[xiii.] \textbf{Restocking intervention:} Restocking was implemented as a one-time addition of susceptible farms after scheduled culling on the candidate restocking date. Two formulations were considered. In the original formulation, a fixed fraction of the baseline county--production population was added to the susceptible compartment: 
   
   \[ A^{\mathrm{original}}_{c,p}(t_r) = \left\lfloor f_{\mathrm{restock}}N_{c,p} \right\rfloor, \]
   
   \[ S_{c,p}(t_r) \leftarrow S_{c,p}(t_r) + A^{\mathrm{original}}_{c,p}(t_r), \] 
   
   where $t_r$ is the candidate restocking date and $f_{\mathrm{restock}}$ is the restocking fraction. Because this formulation did not account for farms that remained present within the county-production stratum at $t_r$, it could increase the county--production population above its baseline capacity. 
   
   The primary restocking analysis therefore used a separate capacity-based formulation. Immediately before restocking, the current number of farms was calculated as 
   
   \[ N^{\mathrm{current}}_{c,p}(t_r) = S_{c,p}(t_r) + E_{c,p}(t_r) + I_{c,p}(t_r) + R_{c,p}(t_r). \] 
   
   The available capacity was 
   
   \[ G_{c,p}(t_r) = \max \left\{ N_{c,p} - N^{\mathrm{current}}_{c,p}(t_r), 0 \right\}. \] 
   
   The number of farms reintroduced under the capacity-based formulation was 
   
   \[ A^{\mathrm{capacity}}_{c,p}(t_r) = \left\lfloor f_{\mathrm{restock}} G_{c,p}(t_r) \right\rfloor. \] 
   
   Restocked farms were assumed to be susceptible and uninfected, and were added only to the susceptible compartment:
   
   \[ S_{c,p}(t_r) \leftarrow S_{c,p}(t_r) + A^{\mathrm{capacity}}_{c,p}(t_r). \]
   
   Under this formulation, $f_{\mathrm{restock}}=0.20$ represented the replacement of 20\% of the available empty capacity rather than the addition of 20\% of the baseline population. The capacity-based formulation ensured that the number of farms in each county--production stratum did not exceed its baseline capacity.

\end{itemize}

\subsection{Scenario analyses}

\subsubsection{Preventive-culling impact} The effect of preventive culling was first evaluated by comparing two scenarios: one including all recorded preventive culling events and one excluding preventive culling. Both scenarios retained the recorded reactive culling schedules and used identical transmission parameters, movement data, spatial connectivity, confinement assumptions, initial conditions, and run-specific random seeds. 

For stochastic run $r$, the cumulative infectious burden was calculated as 
\[ B_r = \sum_t \sum_c \sum_p I_{c,p,r}(t). \]

Because this outcome sums the daily number of infectious farms over the simulation period, it is expressed in infectious-farm-days rather than as a count of distinct outbreaks. 

The mean cumulative burden averted by preventive culling was calculated as 
\[ A_{\mathrm{burden}} = \overline{B}_{\mathrm{without}} - \overline{B}_{\mathrm{with}}, \] 

\noindent where $\overline{B}_{\mathrm{without}}$ and $\overline{B}_{\mathrm{with}}$ denote the ensemble mean cumulative burdens across all simulation runs without and with preventive culling, respectively. The relative reduction was calculated separately for Broiler-2, organic duck, and other production systems using

\[ \mathrm{Reduction}\;(\%) = 100 \times \frac{ \overline{B}_{\mathrm{without}} - \overline{B}_{\mathrm{with}} }{ \overline{B}_{\mathrm{without}} }. \]

A supplementary production-targeted analysis compared three preventive culling policies: no preventive culling, duck-only preventive culling, and chicken-only preventive culling. Corresponding stochastic runs used identical random seeds, allowing run-level burden differences to be evaluated as paired simulation outcomes. For each targeted policy $q$, the paired reduction in run $r$ was

\[ D_{r,q} = B_{r,\mathrm{none}} - B_{r,q}. \]

\noindent Positive values of $D_{r,q}$ indicated that the targeted policy reduced cumulative infectious burden relative to no preventive culling. The mean paired reduction and its approximate 95\% interval were calculated as 

\[ \overline{D}_q \pm 1.96 \frac{s_{D_q}}{\sqrt{R}}, \]

\noindent where $s_{D_q}$ is the standard deviation of the paired differences and $R$ is the number of simulation runs. The proportion of paired runs in which the targeted policy produced a higher burden than no preventive culling was also reported.

\subsubsection{County-level and production-specific burden} 
County-level cumulative infectious burden was calculated for each stochastic run by summing infectious farms over production classes and simulation days: 
\[ B_{c,r} = \sum_t \sum_p I_{c,p,r}(t). \] 

The mean county-level burden was then calculated across runs: 

\[ \overline{B}_c = \frac{1}{R} \sum_{r=1}^{R} B_{c,r}. \] 

To examine production-specific contributions to county-level burden, the corresponding outcome was calculated separately for each county--production stratum: 

\[ B_{c,p,r} = \sum_t I_{c,p,r}(t), \] 
\[ \overline{B}_{c,p} = \frac{1}{R} \sum_{r=1}^{R} B_{c,p,r}. \]

County burdens were ranked from highest to lowest to identify geographic concentrations of simulated infection. These outcomes were interpreted as time-integrated infectious-farm-days and not as counts of distinct infected premises. 

\subsubsection{Five-week spatial burden} 
Spatial epidemic dynamics were summarised over consecutive 35-day periods. For county $c$, production class $p$, and five-week period $w$, cumulative burden was calculated as 
\[ W_{c,p,w} = \sum_{t\in w} \overline{I}_{c,p}(t), \] 

where \[ \overline{I}_{c,p}(t) = \frac{1}{R} \sum_{r=1}^{R} I_{c,p,r}(t). \] 

The overall five-week burden in a county was obtained by summing over production classes as 
\[ W_{c,w} = \sum_p W_{c,p,w}. \] 

The periods were 22 December 2025--25 January 2026, 26 January--1 March, 2 March--5 April, 6 April--10 May, 11 May--14 June, and 15--30 June 2026. The final interval was shorter than 35 days because the simulation ended on 30 June 2026. Maps were produced for all production systems combined and separately for Broiler-2, organic duck, and other farms. Mapped values represented cumulative infectious-farm-days within each period.

\subsubsection{Confinement analyses \label{sec:conf_analy}} The effect of confinement was evaluated through three related analyses: confinement timing, environmental-transmission strength, and confinement scope. These analyses varied one component at a time while retaining the remaining model assumptions.

For the timing analysis, confinement start dates of 31 December 2025, 7 January 2026, 14 January 2026, and 14 February 2026 were evaluated. For each date, the ensemble mean epidemic trajectory, peak mean number of infectious farms, date of the epidemic peak, and mean cumulative infectious burden were calculated. 

The sensitivity of the epidemic to environmental transmission was evaluated by multiplying all production-specific environmental transmission coefficients by 
\[ q_{\mathrm{env}} \in \{0.1,0.5,1.0\}. \] 

For each multiplier, \[ \beta^{\mathrm{env},q}_p = q_{\mathrm{env}} \beta^{\mathrm{env}}_p. \]

The confinement start date and remaining transmission parameters were held fixed during this analysis. Confinement scope was evaluated by changing the post-confinement environmental exposure multipliers. The principal configurations included the baseline Broiler-2 and organic-duck confinement, 

\[ \widetilde{m}^{\mathrm{exp}}_p = (0,0,1), \]

Broiler-2-only confinement, 
\[ \widetilde{m}^{\mathrm{exp}}_p = (0,1,1), \] 

organic-duck-only confinement,

\[ \widetilde{m}^{\mathrm{exp}}_p = (1,0,1), \]

and full confinement of all production classes, 
\[ \widetilde{m}^{\mathrm{exp}}_p = (0,0,0), \] 

\noindent with values reported in the order of $(B2,D,O)$. A value of zero indicated complete interruption of the modelled environmental-exposure pathway for the corresponding production class, whereas a value of one indicated no reduction in environmental exposure. Environmental shedding and movement-mediated transmission multipliers remained unchanged unless otherwise specified.

\subsubsection{Restocking timing and rebound analysis} Candidate restocking dates were evaluated at seven-day intervals from 1 February to 28 June 2026. For each candidate date, 200 stochastic simulations were conducted under the all-recorded preventive culling scenario, with confinement beginning on 14 February 2026. Restocking was implemented once, after the epidemiological transitions and scheduled culling events on the candidate date. The primary analysis used the capacity-based formulation described in the restocking-intervention component of the modelling framework, with 

\[ f_{\mathrm{restock}}=0.20. \]

\noindent Thus, 20\% of the available empty capacity in each county--production stratum was restored. Restocked farms were assumed to be susceptible and uninfected at reintroduction. 

For each stochastic run $r$, the total number of infectious farms on day $t$ was calculated by summing over counties and production classes:

\[ I_r(t) = \sum_c \sum_p I_{c,p,r}(t). \]

For candidate restocking date $t_r$, the infectious count on that date was used as the run-specific baseline. A run was classified as exhibiting an epidemic rebound if the maximum infectious count observed from the restocking date to the end of the simulation exceeded this baseline by more than five infectious farms: 

\[ Z_r(t_r) = \mathbf{1} \left[ \max_{t\geq t_r} I_r(t) > I_r(t_r)+\tau \right], \]

where $\tau=5$ infectious farms and $\mathbf{1}[\cdot]$ is the indicator function. 

The rebound probability for candidate date $t_r$ was estimated as 
\[ \widehat{P}_{\mathrm{rebound}}(t_r) = \frac{1}{R} \sum_{r=1}^{R} Z_r(t_r), \]

where $R=200$ is the number of stochastic simulation runs. Candidate dates were classified as 

\[ \mathrm{Classification}(t_r) = \begin{cases} \mathrm{SAFE}, & \widehat{P}_{\mathrm{rebound}}(t_r) \leq \alpha,\\[4pt] \mathrm{RISKY}, & \widehat{P}_{\mathrm{rebound}}(t_r) > \alpha, \end{cases} \]

where the prespecified safety threshold was $\alpha=0.20$. 

Additional outcomes included the median number of infectious farms on the restocking date, the median maximum post-restocking infectious count, and the 97.5th percentile of the maximum post-restocking infectious count across stochastic runs. The latter was interpreted as an upper simulation quantile rather than a confidence limit.

The sensitivity of rebound risk to restocking intensity was evaluated by varying $f_{\mathrm{restock}}$ from 0.10 to 1.00 in increments of 0.10, while fixing the candidate restocking date at 15 March 2026. For each fraction, 200 stochastic simulations were conducted, and the rebound probability, median epidemic peak, and 97.5th percentile of the epidemic peak were calculated.

As a formulation sensitivity analysis, the capacity-based formulation was compared with the original formulation described above. Both formulations were evaluated using a candidate restocking date of 15 March 2026, a restocking fraction of 0.20, identical model parameters and intervention settings, and matching run-specific random seeds. The formulations were compared using mean cumulative infectious-farm-days and estimated rebound probability.

\subsection{Parameter values} 
The main parameter values used in the simulations are summarised in Table~\ref{tab:model_parameters}. Production-specific parameters are reported in the order $(B2,D,O)$, corresponding to Broiler-2 chicken farms, organic duck farms, and all other farms, respectively. The baseline post-confinement environmental-exposure multipliers were $(0,0,1)$. Thus, confinement interrupted the modelled environmental exposure pathway for Broiler-2 and organic duck farms, while environmental exposure remained unchanged for farms in the Other production class. Environmental shedding and movement-mediated transmission were not modified by confinement in the baseline scenario. The transmission coefficients were treated as scenario parameters rather than statistically estimated quantities.

\begin{sidewaystable}[p] 
\centering 
\caption{ Model parameters used in the stochastic county--production SEIR metapopulation model. Production-specific values are reported in the order $(B2,D,O)$. Transmission coefficients are simulation-scale parameters whose effective units depend on their corresponding model inputs. } \label{tab:model_parameters}
\renewcommand{\arraystretch}{1.25} \setlength{\tabcolsep}{8pt} \small \begin{tabular}{ >{\raggedright\arraybackslash}p{3.cm} >{\raggedright\arraybackslash}p{7.cm} >{\centering\arraybackslash}p{4.0cm} >{\raggedright\arraybackslash}p{6.cm} } \toprule \textbf{Parameter} & \textbf{Description} & \textbf{Value} & \textbf{Unit / interpretation} \\ \midrule $\beta^{\mathrm{within}}_p$ & Local within-county transmission coefficient & $(0.30,\,0.20,\,0.15)$ & Simulation-scale coefficient \\ $\beta^{\mathrm{env}}_p$ & Environmental transmission coefficient & $(0.10,\,0.08,\,0.15)$ & Per contamination unit \\ $\kappa^{\mathrm{env}}_p$ & Environmental shedding coefficient & $(0.30,\,0.25,\,0.40)$ & Contamination units per infectious farm-day \\ $\beta^{\mathrm{move}}_p$ & Movement-mediated transmission coefficient & $(4,\,3,\,2)\times10^{-4}$ & Per movement-volume unit \\ $\beta^{\mathrm{spatial}}_p$ & Spatial transmission coefficient & $(0.05,\,0.04,\,0.03)$ & Per spatially weighted infectious farm \\ $\sigma$ & Daily exposed-to-infectious transition probability & $1/3$ & Mean latent duration of approximately 3 days \\ $\gamma$ & Daily infectious-to-removed transition probability & $1/7$ & Mean infectious duration of approximately 7 days \\ $\delta_{\mathrm{env}}$ & Daily environmental decay proportion & $0.20$ & 20\% of environmental contamination removed per day \\ $d_0$ & Spatial decay parameter & $2000$ & m \\ $\varepsilon$ & Small denominator constant & $10^{-6}$ & Prevents division by zero \\ $h_i$ & Farm-level hazard-risk-zone multiplier & $1.0$ or $1.5$ & Outside or inside the designated hazard-risk zone \\ $\widetilde{m}^{\mathrm{exp}}_p$ & Post-confinement environmental-exposure multiplier & $(0,\,0,\,1)$ & Environmental exposure interrupted for $B2$ and $D$ \\ $\widetilde{m}^{\mathrm{shed}}_p$ & Post-confinement environmental-shedding multiplier & $(1,\,1,\,1)$ & No reduction in the baseline configuration \\ $\widetilde{m}^{\mathrm{mov}}_p$ & Post-confinement movement-transmission multiplier & $(1,\,1,\,1)$ & No reduction in the baseline configuration \\ $f_{\mathrm{restock}}$ & Fraction of available capacity restored & $0.20$ & Primary capacity-based restocking scenario \\ $\alpha$ & Safety threshold for rebound probability & $0.20$ & Probability \\ $\tau$ & Rebound tolerance & $5$ & Infectious farms \\ $R$ & Number of stochastic simulation runs & $200$ & Unless otherwise specified \\ 
\bottomrule
\end{tabular}
\end{sidewaystable}

\section{Results}

\subsection{Temporal epidemic dynamics}\label{sec:temporaldynamics}

\begin{figure}[pos=htbp]
\centering
\includegraphics[ width=0.7\textwidth ]{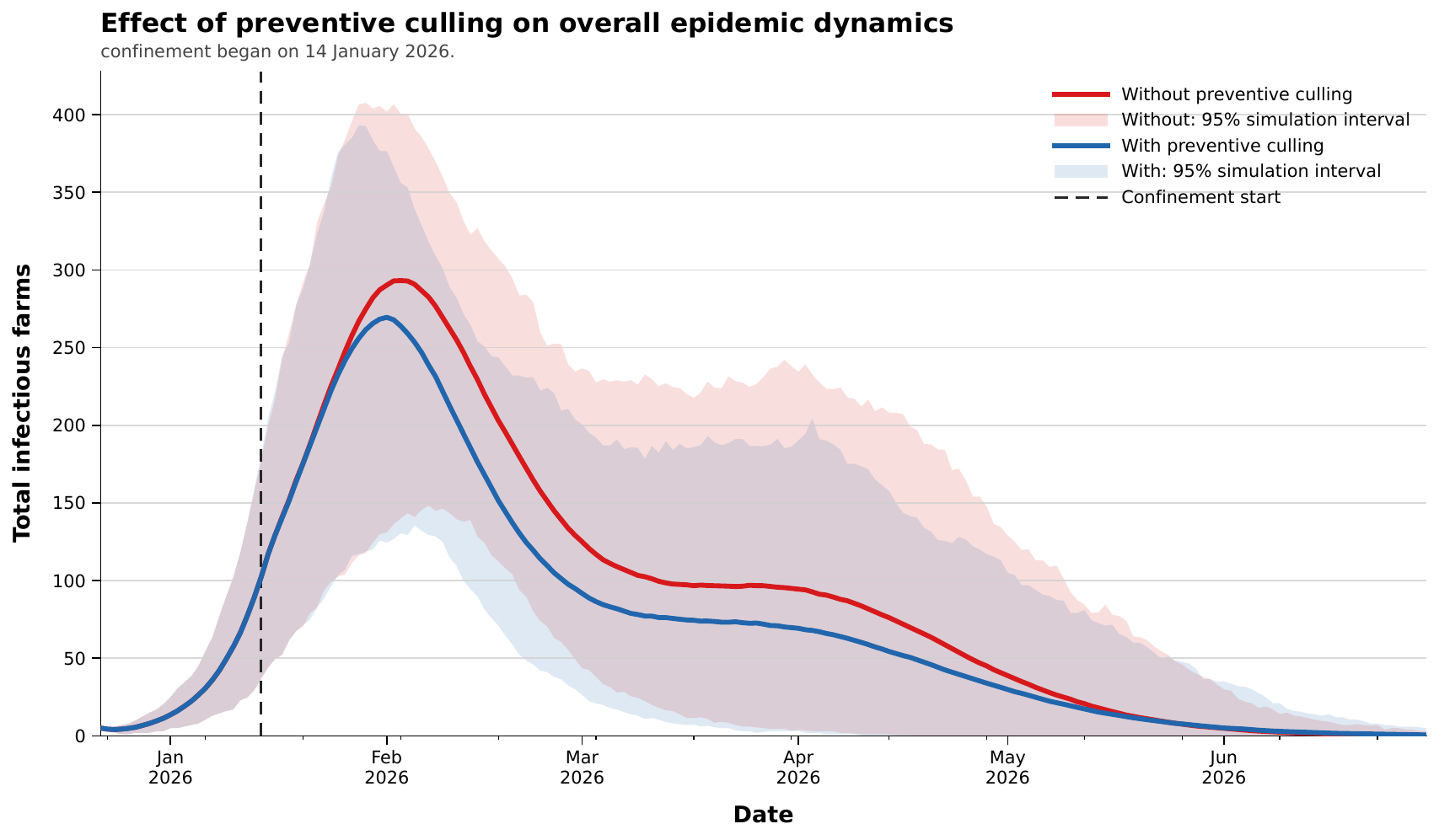} 
\caption{ Simulated overall HPAI epidemic trajectories with (blue) and without (red) preventive culling. Solid lines represent the ensemble-mean number of infectious farms, and shaded areas represent the 2.5th and 97.5th percentiles across stochastic simulation runs. The vertical dashed black line indicates the start of confinement. } 
\label{fig:overall_epidemic} 
\end{figure}

Figure~\ref{fig:overall_epidemic} shows the ensemble-mean number of infectious farms from 22 December 2025 to 30 June 2026 under scenarios with and without all recorded preventive culling. In both scenarios, the number of infectious farms remained relatively low during the initial simulation period before increasing rapidly in January 2026. The epidemic reached its principal peak in early February, after the start of confinement on 14 January 2026. 

Following the principal peak, the mean number of infectious farms declined during February and March. However, the decline was not monotonic, a plateau and modest secondary increase were evident during March and early April, followed by a sustained decline towards near-zero infection levels by June 2026. The 95\% simulation intervals were narrow during the initial growth phase, widened around the epidemic peak and the March--April plateau, and narrowed again as the epidemic approached resolution. This pattern indicated greater between-run stochastic variation during periods of sustained transmission. 

As observed in Figure~\ref{fig:overall_epidemic}, the mean trajectory under preventive culling remained below the trajectory without preventive culling for much of the epidemic period. The separation was most apparent around and after the principal epidemic peak. A formal comparison of cumulative infectious-farm-days under the two scenarios is presented in Section~\ref{sec:preventive_culling}. 

\begin{figure}[pos=htbp] 
\centering 
\includegraphics[ width=0.7\textwidth ]{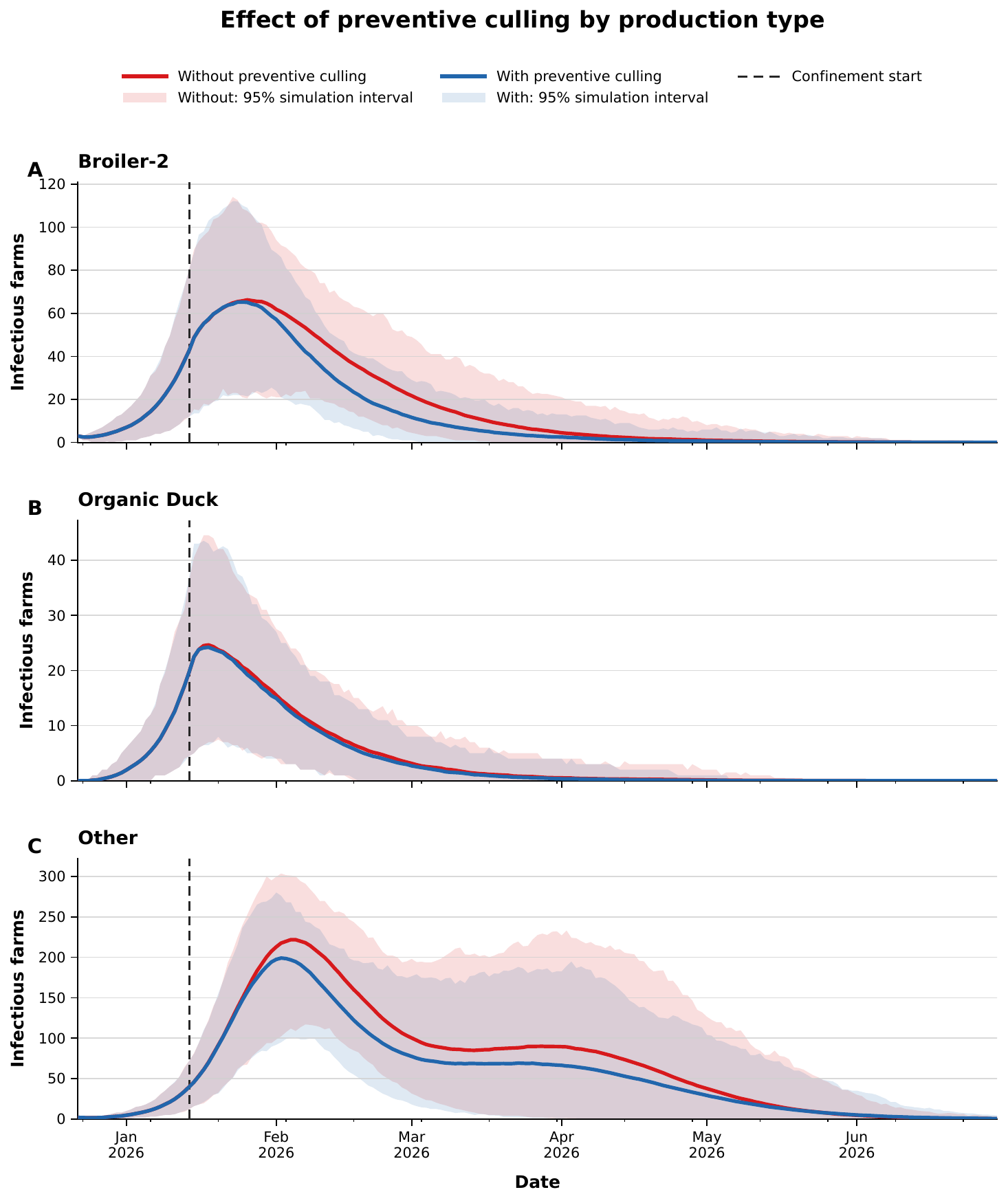}
\caption{ Simulated HPAI epidemic trajectories by production class under scenarios with (blue) and without (red) preventive culling. Panels show Broiler-2 farms (A), organic duck farms (B), and Other production systems (C). Solid lines represent ensemble means, and shaded areas represent the 2.5th and 97.5th percentiles across stochastic simulation runs. The vertical dashed black lines indicate the start of confinement. } 
\label{fig:epidemic_by_production}
\end{figure}

Figure~\ref{fig:epidemic_by_production} decomposes the overall epidemic trajectory by production class. The Other production class contributed the largest number of infectious farms and displayed the most pronounced multi-phase trajectory. Its mean trajectory reached a principal peak in early February, declined temporarily, and subsequently exhibited a broader secondary increase during March and early April before approaching zero. 

Broiler-2 farms exhibited a more concentrated epidemic trajectory, with a principal peak in late January or early February followed by a comparatively steady decline. Organic duck farms experienced the smallest epidemic burden of the three production classes and returned towards low infection levels earlier than the other classes. As such, the production-specific trajectories indicate that the March--April plateau in the overall epidemic was driven primarily by infections in the Other production class rather than by a uniform secondary increase across all production systems. As seen in Figure~\ref{fig:epidemic_by_production}, preventive culling reduced the ensemble-mean trajectory in each production class, although the magnitude of the separation differed among classes. The difference was most apparent for Broiler-2 and Other farms and was comparatively small for organic duck farms.

\subsection{County-level burden and spatiotemporal dynamics} \label{sec:county_spatial_results} 

\begin{figure}[pos=htbp] 
\centering 
\begin{subfigure}[t]{0.48\textwidth}
\centering 
\includegraphics[ width=\linewidth ]{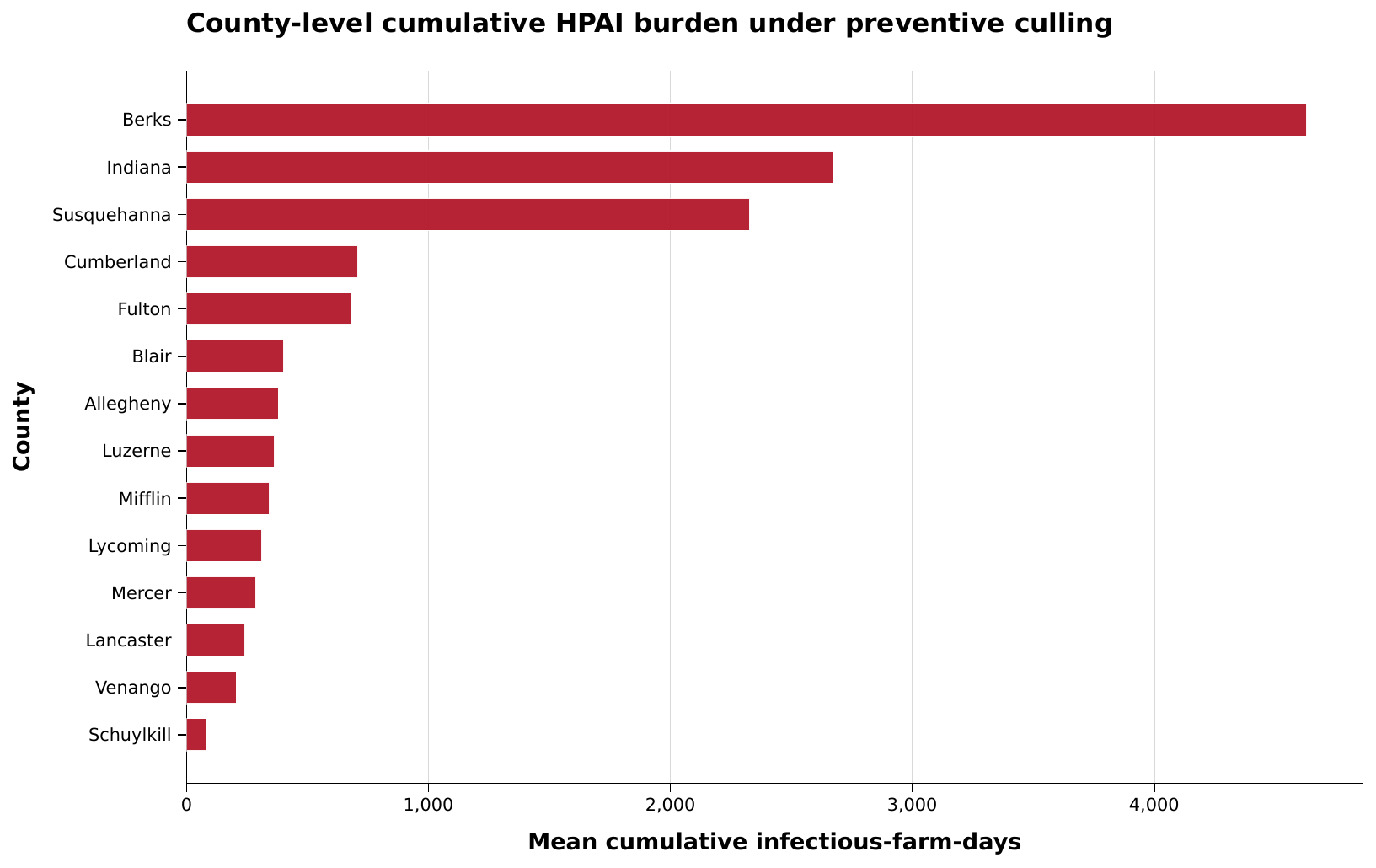} 
\caption{ Overall cumulative infectious burden by county. } 
\label{fig:county_overall_burden} \end{subfigure}
\hfill 
\begin{subfigure}[t]{0.50\textwidth} 
\centering \includegraphics[ width=\linewidth ]{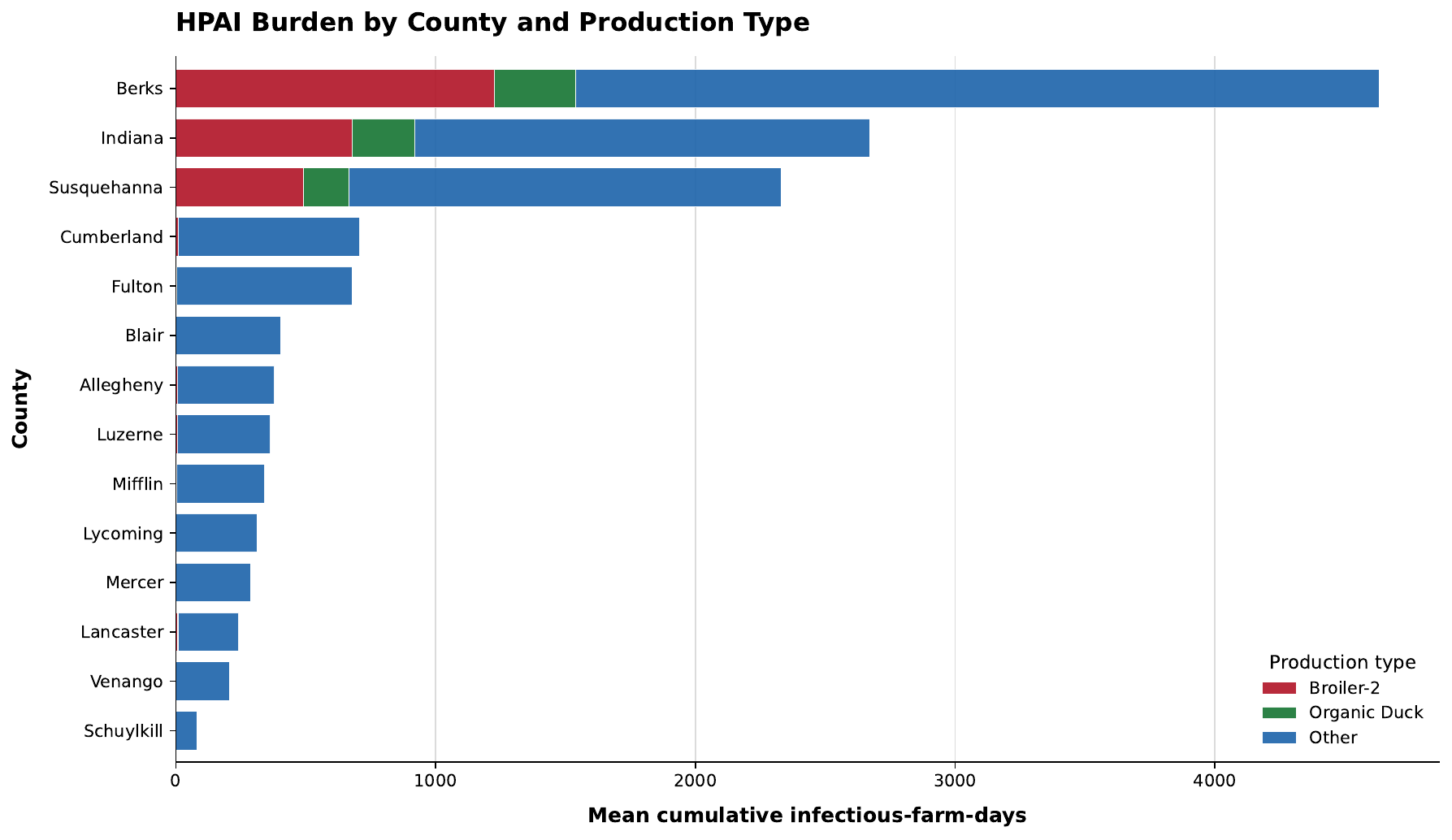} 
\caption{ County-level burden partitioned by production class. } 
\label{fig:county_production_burden} \end{subfigure} 
\caption{ Mean cumulative HPAI burden by county under the preventive-culling scenario. Panel~(a) shows the overall cumulative burden, summed across production classes and the full simulation period. Panel~(b) partitions the county-level burden into contributions from Broiler-2 (red), organic duck (green), and Other (blue) production systems. Values represent ensemble mean infectious-farm-days across stochastic simulation runs. } 
\label{fig:county_burden} 
\end{figure}

Figure~\ref{fig:county_burden} presents the cumulative HPAI burden across counties under the preventive culling scenario, both overall and by production class. The overall burden was expressed as the mean cumulative number of infectious farm-days across stochastic simulation runs. Considerable geographic heterogeneity was evident, with the simulated epidemic burden concentrated in a relatively small number of counties, while most counties experienced substantially lower levels of sustained transmission. 

As shown in Figure~\ref{fig:county_overall_burden}, Berks County had the largest mean cumulative burden, at approximately 4,800--5,000 infectious-farm-days. This was nearly twice the burden observed in Indiana, the second-highest-burden county, at approximately 2,700 infectious-farm-days. Susquehanna also contributed substantially to the total epidemic burden, although its cumulative burden remained below those of Berks and Indiana.

The remaining counties experienced comparatively smaller cumulative burdens. The strongly skewed distribution indicates that the regional epidemic was dominated by a limited number of high-burden locations rather than being uniformly distributed across Jolly Island. The high burden in Berks reflects a combination of a greater number of infectious farms and a longer duration of infection within that county. However, county-level cumulative burden alone does not identify the relative contributions of local, environmental, movement-mediated, or spatial transmission. 

Figure~\ref{fig:county_production_burden} partitions the cumulative county burden into the three production classes. Other production systems accounted for the largest share of infectious-farm-days in Berks and in most other counties. Broiler-2 farms made a visible but smaller contribution, particularly in the principal high-burden counties, whereas organic duck farms accounted for a comparatively small proportion of the cumulative burden.

\begin{figure}[pos=htbp]
\centering 
\includegraphics[ width=0.7\textwidth ]{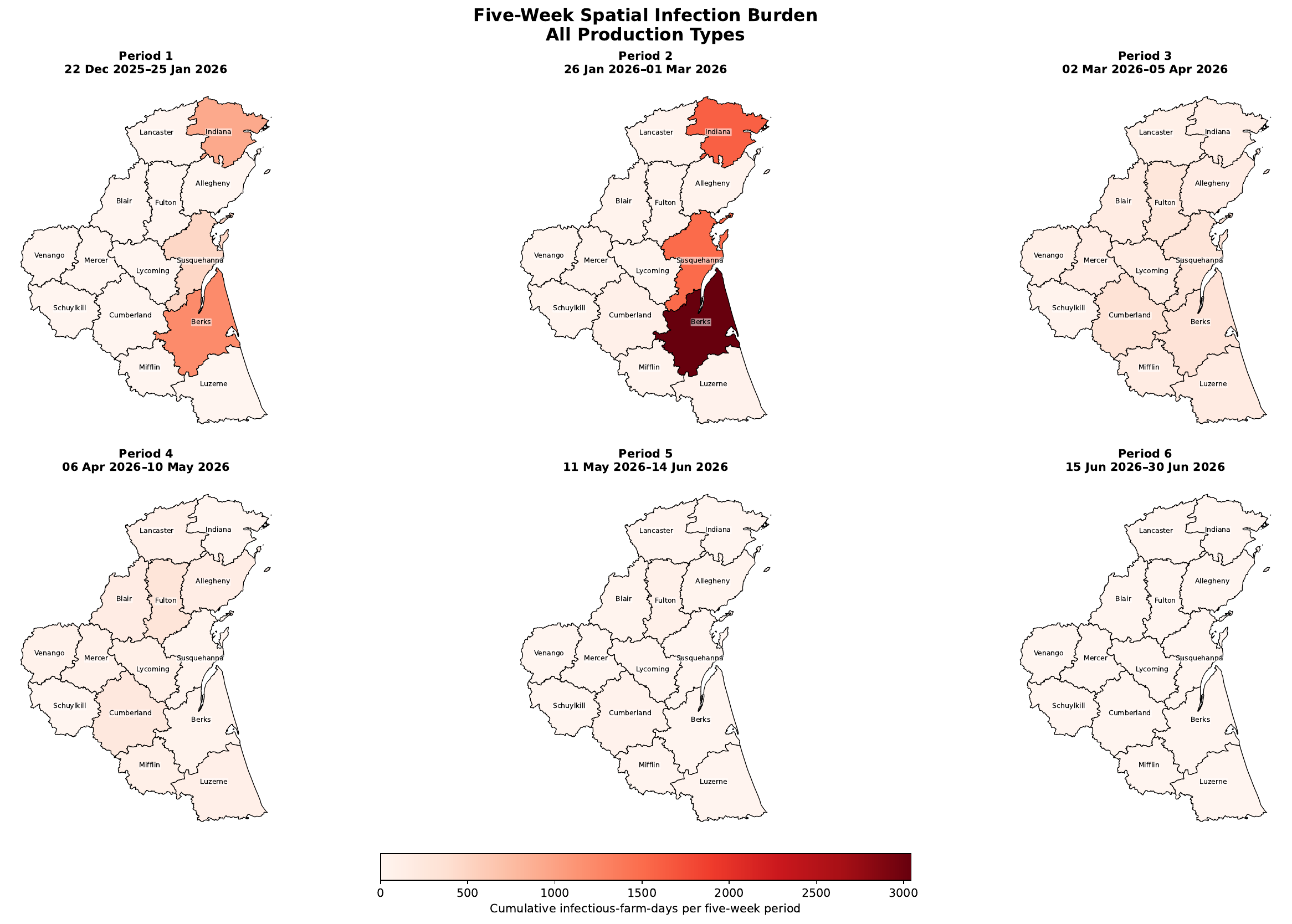} 
\caption{ County-level cumulative HPAI burden across all production classes during consecutive five-week periods. Values represent ensemble mean cumulative infectious-farm-days within each period. The periods were 22 December 2025--25 January 2026, 26 January--1 March 2026, 2 March--5 April 2026, 6 April--10 May 2026, 11 May--14 June 2026, and 15--30 June 2026. The final period was shorter than five weeks because the simulation ended on 30 June 2026. } 
\label{fig:spatial_overall} 
\end{figure}

\begin{figure}[pos=htbp]
\centering 
\includegraphics[ width=0.7\textwidth ]{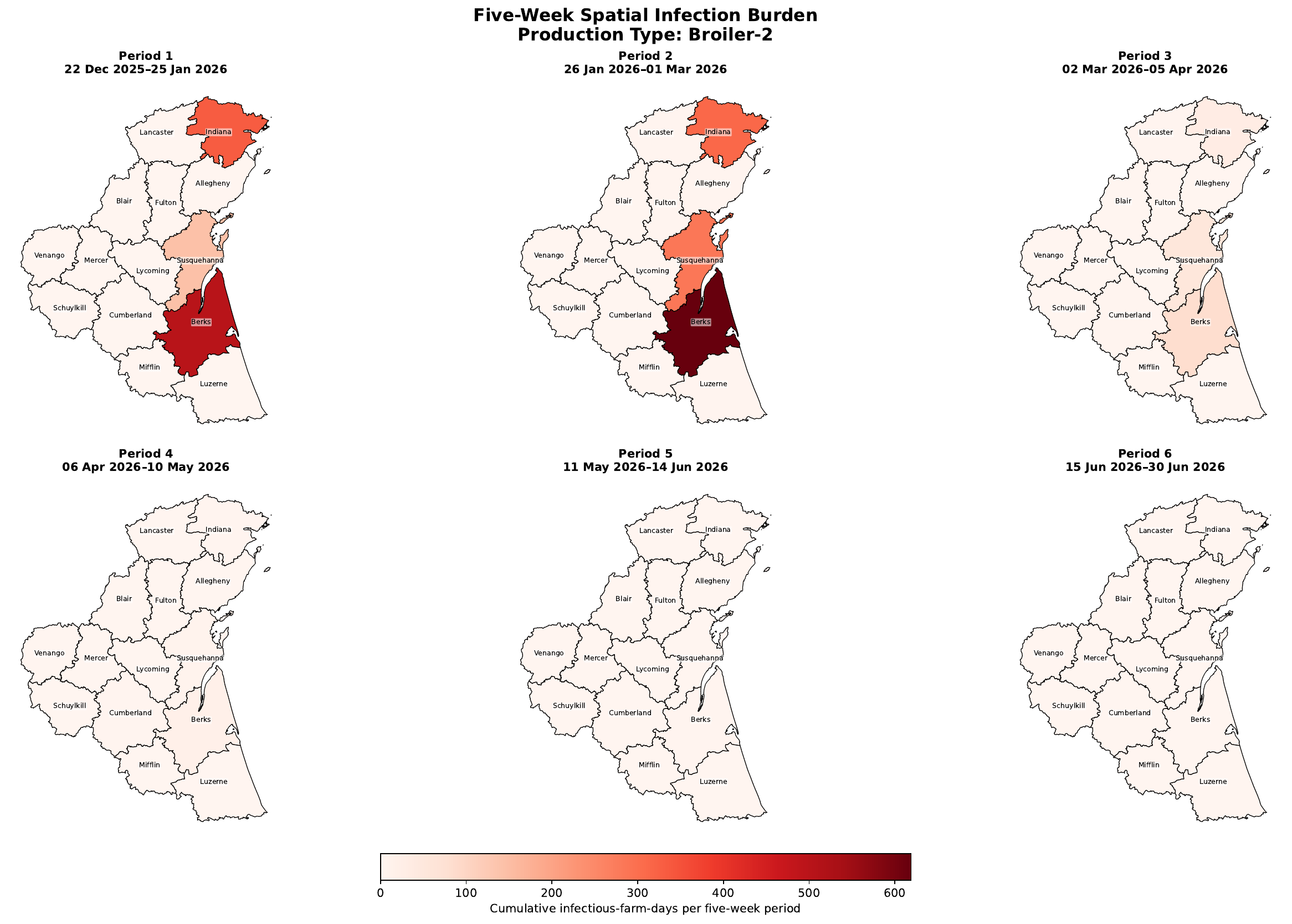}
\caption{ County-level cumulative HPAI burden among Broiler-2 farms during consecutive five-week periods. Values represent ensemble mean cumulative infectious-farm-days within each period. The final period, 15--30 June 2026, was shorter than five weeks. }
\label{fig:spatial_broiler} 
\end{figure}

\begin{figure}[pos=htbp] 
\centering 
\includegraphics[ width=0.7\textwidth ]{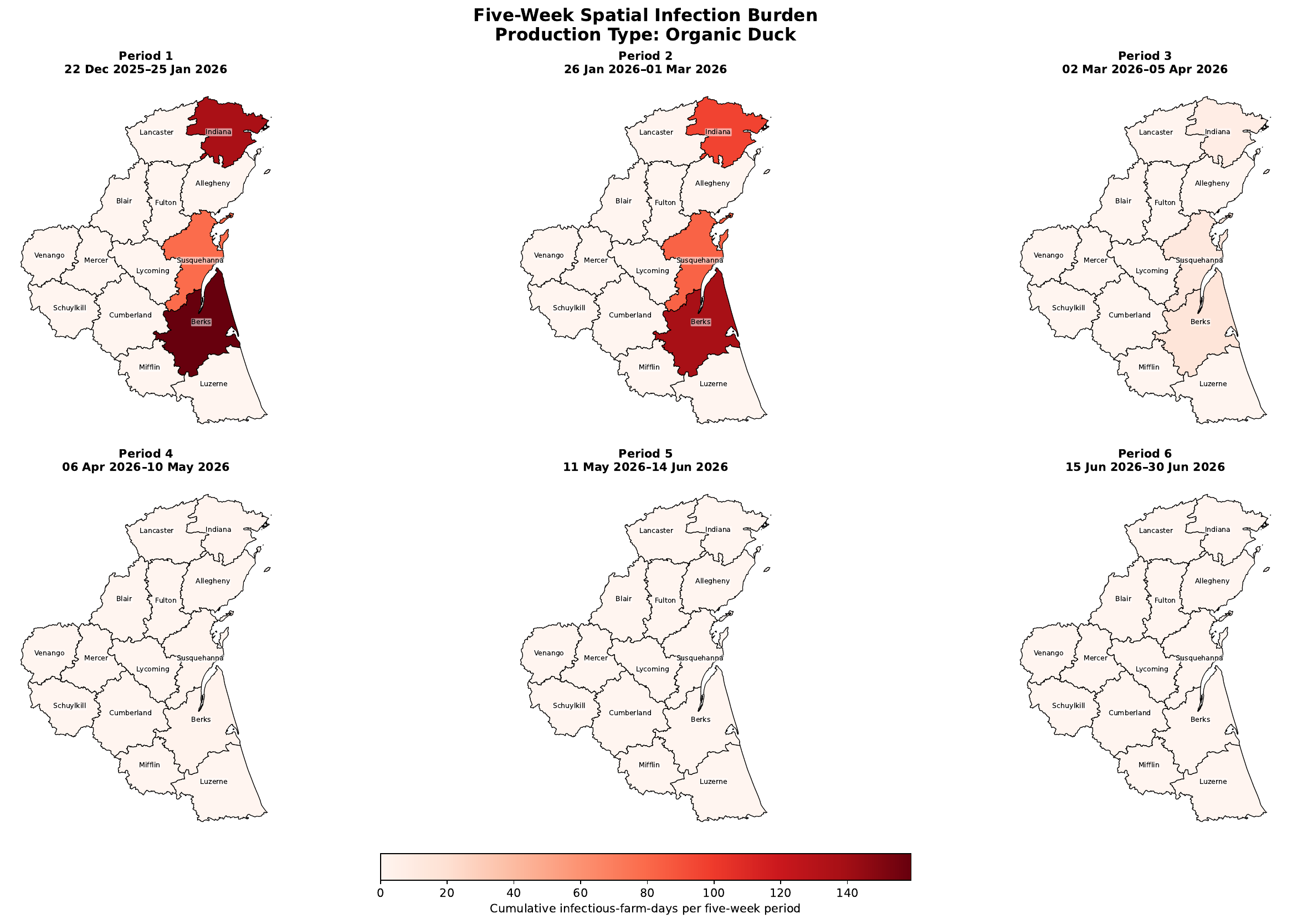} 
\caption{ County-level cumulative HPAI burden among organic duck farms during consecutive five-week periods. Values represent ensemble mean cumulative infectious-farm-days within each period. The final period, 15--30 June 2026, was shorter than five weeks. } 
\label{fig:spatial_duck} 
\end{figure}

\begin{figure}[pos=htbp] \centering
\includegraphics[ width=0.7\textwidth ]{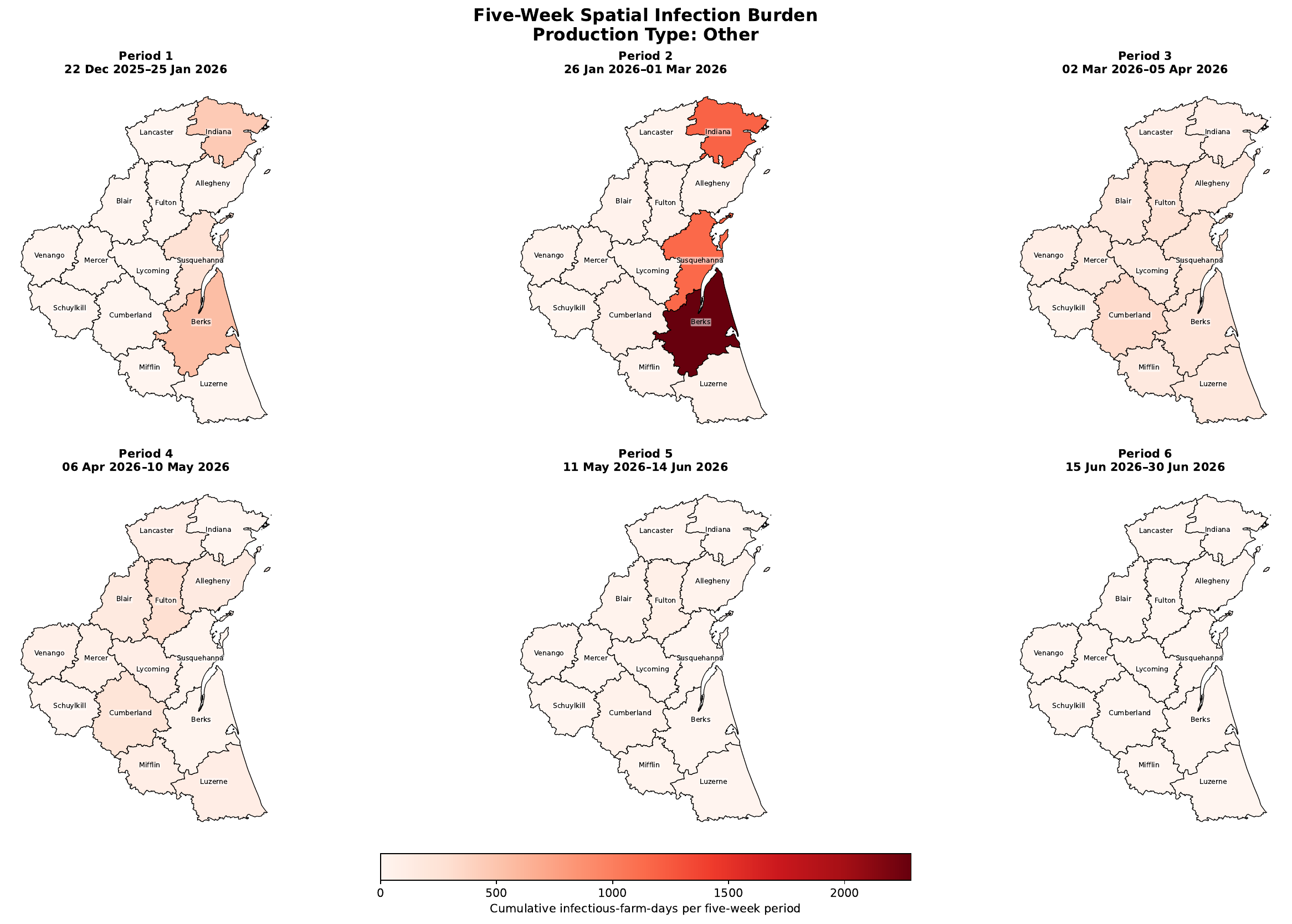} 
\caption{ County-level cumulative HPAI burden among Other production systems during consecutive five-week periods. Values represent ensemble mean cumulative infectious-farm-days within each period. The final period, 15--30 June 2026, was shorter than five weeks. }
\label{fig:spatial_other} 
\end{figure}

In several of the lower-burden counties, most of the simulated burden was also associated with Other production systems. This production-specific composition was consistent with the temporal results, in which the Other production class generated the highest and most persistent infectious trajectory. Nevertheless, the relative contribution of each production class varied among counties, indicating that geographic burden reflected both location and local production composition. The county-level burden accumulated over the complete simulation period, whereas the spatiotemporal maps in Figures~\ref{fig:spatial_overall}--\ref{fig:spatial_other} show how this burden developed over consecutive periods. Figure~\ref{fig:spatial_overall} shows the combined burden across all production classes. During the first period, from 22 December 2025 to 25 January 2026, infection was concentrated principally in Berks and Indiana. During the second period, from 26 January to 1 March 2026, epidemic intensity increased within these counties and expanded into neighbouring areas, including Susquehanna and Fulton. 

The production-specific maps demonstrate that Broiler-2, organic duck, and Other farms experienced infection within a broadly similar geographic footprint, although the intensity and persistence of burden differed among production classes. In Figure~\ref{fig:spatial_broiler}, we observed that Broiler-2 burden was concentrated principally in the high-burden counties and declined after the main epidemic period. Organic duck farms experienced the lowest spatial burden of the three production classes. Their burden was restricted primarily to the main high-burden counties and declined towards negligible levels earlier than the burden in Broiler-2 and Other farms, see Figure~\ref{fig:spatial_duck}. As seen in Figure~\ref{fig:spatial_other}, Other farms generated the highest spatial burden and the most persistent pattern across the simulation period. This production class accounted for most of the burden visible in the combined map, particularly in Berks, Indiana, and Susquehanna. 

The maps show a geographically contiguous cluster of high-burden counties in the north-central part of the study region. This clustering was consistent with the spatial and movement-mediated transmission processes represented in the model. However, because local, environmental, movement-mediated, and distance-dependent transmission operated simultaneously, the maps alone could not determine the relative contribution of each pathway to the burden observed in an individual county.

The geographic footprint was broadly consistent across the combined and production-specific maps, while the burden magnitude differed substantially among production classes. The combined and Other-production maps reached approximately 2,000--2,500 infectious-farm-days per period in the principal hotspot counties. By comparison, the corresponding scales reached approximately 400 infectious-farm-days for Broiler-2 farms and approximately 140 infectious-farm-days for organic duck farms. Because each map used a production-specific colour scale, colours should be interpreted within each figure rather than compared directly across production classes.

During the third and subsequent periods, burden declined within the original high-burden counties, and no comparably intense new geographic foci emerged. By the final simulation periods, the burden in most counties had approached negligible levels. This pattern indicates that the simulated epidemic contracted within its established geographic footprint rather than continuing to expand into previously low-burden counties. The decline in spatial burden coincided with the downward phase of the regional epidemic. However, the reduction cannot be attributed to a single control measure because confinement, reactive and preventive culling, environmental decay, disease progression, and susceptible depletion operated concurrently. Taken together, the county-level burden plots and spatiotemporal maps demonstrate that the simulated outbreak was characterised by a stable group of high-burden counties and substantial production-specific differences in the magnitude and persistence of infection.

\subsection{Effect of preventive culling under different confinement scenarios} \label{sec:preventive_culling} 

Here we evaluate two related control dimensions: the production class targeted by preventive culling and the production classes protected by confinement. Preventive culling scenarios included no preventive culling, duck-targeted preventive culling, chicken-targeted preventive culling, and, in the principal analysis, all recorded preventive culling. Confinement scope was varied by modifying the production classes for which environmental exposure was interrupted after the start date of confinement.

The baseline confinement configuration interrupted environmental exposure for both Broiler-2 and organic duck farms, while the Other production class remained environmentally exposed. Additional scenarios considered Broiler-2-only and organic-duck-only confinement. Within each confinement configuration, no preventive culling, duck-targeted preventive culling, and chicken-targeted preventive culling were compared using matching stochastic runs and otherwise identical model assumptions.

As shown in Figures~\ref{fig:overall_epidemic} and \ref{fig:epidemic_by_production}, preventive culling reduced the ensemble-mean infectious trajectory during much of the simulated epidemic. Table~\ref{tab:preventive_culling_by_production} summarises the cumulative burden under the baseline confinement configuration by comparing all recorded preventive culling with no preventive culling. Across all production classes, preventive culling reduced the mean cumulative burden from 16,362.7 to 13,631.9 infectious-farm-days. This corresponded to 2,730.8 infectious-farm-days averted and an overall relative reduction of 16.7\%. The largest proportional reduction occurred among Broiler-2 farms, for which cumulative burden decreased by 18.2\%. The reduction was 17.0\% among Other production systems and 5.1\% among organic duck farms. 

\begin{table}[pos=htbp]
\centering 
\caption{ Mean cumulative infectious burden with and without all recorded preventive culling, overall and by production class, across 200 stochastic simulation runs. Burden and averted values are expressed as infectious-farm-days. The relative reduction was calculated against the corresponding scenario without preventive culling. } 
\label{tab:preventive_culling_by_production} 
\begin{tabular}{lrrrr} 
\hline \textbf{Production} & \textbf{Without} & \textbf{With} & \textbf{Averted} & \textbf{Reduction (\%)} \\
\hline
Overall & 16,362.713 & 13,631.883 & 2,730.830 & 16.689 \\
Broiler-2 & 2,974.277 & 2,433.510 & 540.767 & 18.181 \\ 
Organic Duck & 773.437 & 733.727 & 39.710 & 5.134 \\ 
Other & 12,615.000 & 10,464.647 & 2,150.353 & 17.046 \\ 
\hline 
\end{tabular} 
\end{table} 

The absolute reduction in burden was driven primarily by the Other production class, which accounted for 2,150.4 of the 2,730.8 infectious-farm-days averted. Broiler-2 farms contributed a smaller absolute reduction of 540.8 infectious-farm-days but exhibited the largest proportional reduction. Organic duck farms contributed comparatively little to the overall reduction, reflecting both their smaller baseline epidemic burden and the weaker simulated effect of preventive culling within this class. 

\subsubsection{Production-targeted preventive culling under baseline confinement} 

\begin{figure}[pos=htbp] 
\centering 
\includegraphics[ width=0.5\linewidth ]{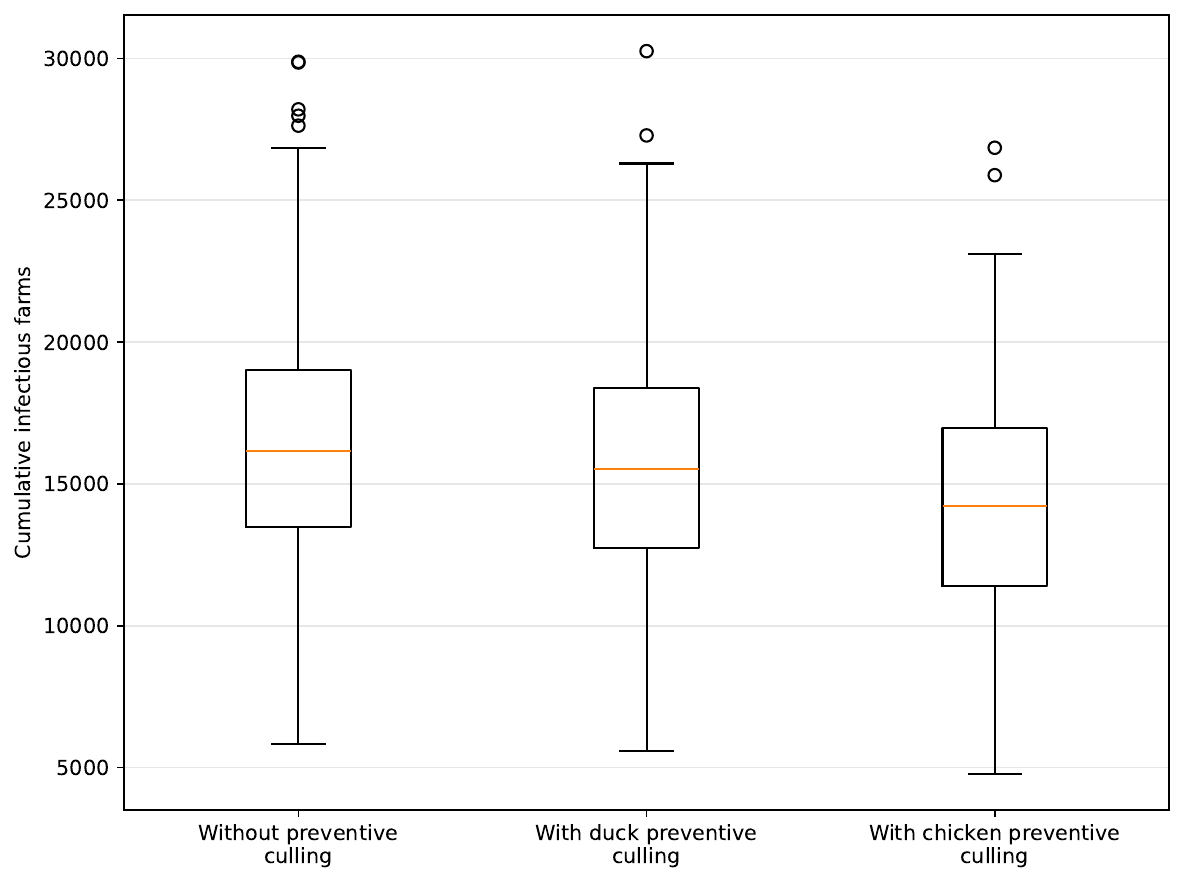} 
\caption{ Boxplots showing the distribution of cumulative infectious burden under no preventive culling, duck-targeted preventive culling, and chicken-targeted preventive culling under the baseline confinement configuration. Confinement began on 14 January 2026 and interrupted environmental exposure for Broiler-2 and organic duck farms. Burden is expressed as cumulative infectious-farm-days across the simulation period. }
\label{fig:preventive_culling_boxplot_baseline}
\end{figure}

A separate targeted-policy analysis compared no preventive culling, duck-targeted preventive culling, and chicken-targeted preventive culling under the baseline confinement configuration. In this configuration, environmental exposure was interrupted for Broiler-2 and organic duck farms from 14 January 2026, while environmental exposure remained unchanged for the Other production class. Figure~\ref{fig:preventive_culling_boxplot_baseline} shows the distributions of cumulative infectious-farm-days under the three targeted policies. The median cumulative burden decreased from approximately 16,100 infectious-farm-days without preventive culling to approximately 15,500 under duck-targeted preventive culling and approximately 14,200 under chicken-targeted preventive culling. 

Chicken-targeted preventive culling produced a larger downward shift in the burden distribution than duck-targeted culling. Its interquartile range was approximately 11,400--17,000 infectious-farm-days, compared with approximately 13,400--19,000 without preventive culling and 12,700--18,400 under duck-targeted preventive culling. The duck-targeted distribution overlapped substantially with the no-preventive-culling distribution, indicating that the effect of duck-targeted culling was small relative to the stochastic variation among simulation runs.

These targeted-policy results should be distinguished from Table~\ref{tab:preventive_culling_by_production}. The table compares all recorded preventive culling with no preventive culling and then partitions the resulting burden by production class. In contrast, Figure~\ref{fig:preventive_culling_boxplot_baseline} compares policies in which preventive culling was restricted according to the species of the recorded farm. 

\subsubsection{Interaction with confinement scope}

\begin{figure}[pos=htbp] 
\centering 
\begin{subfigure}[t]{0.48\textwidth} 
\centering 
\includegraphics[ width=\linewidth ]{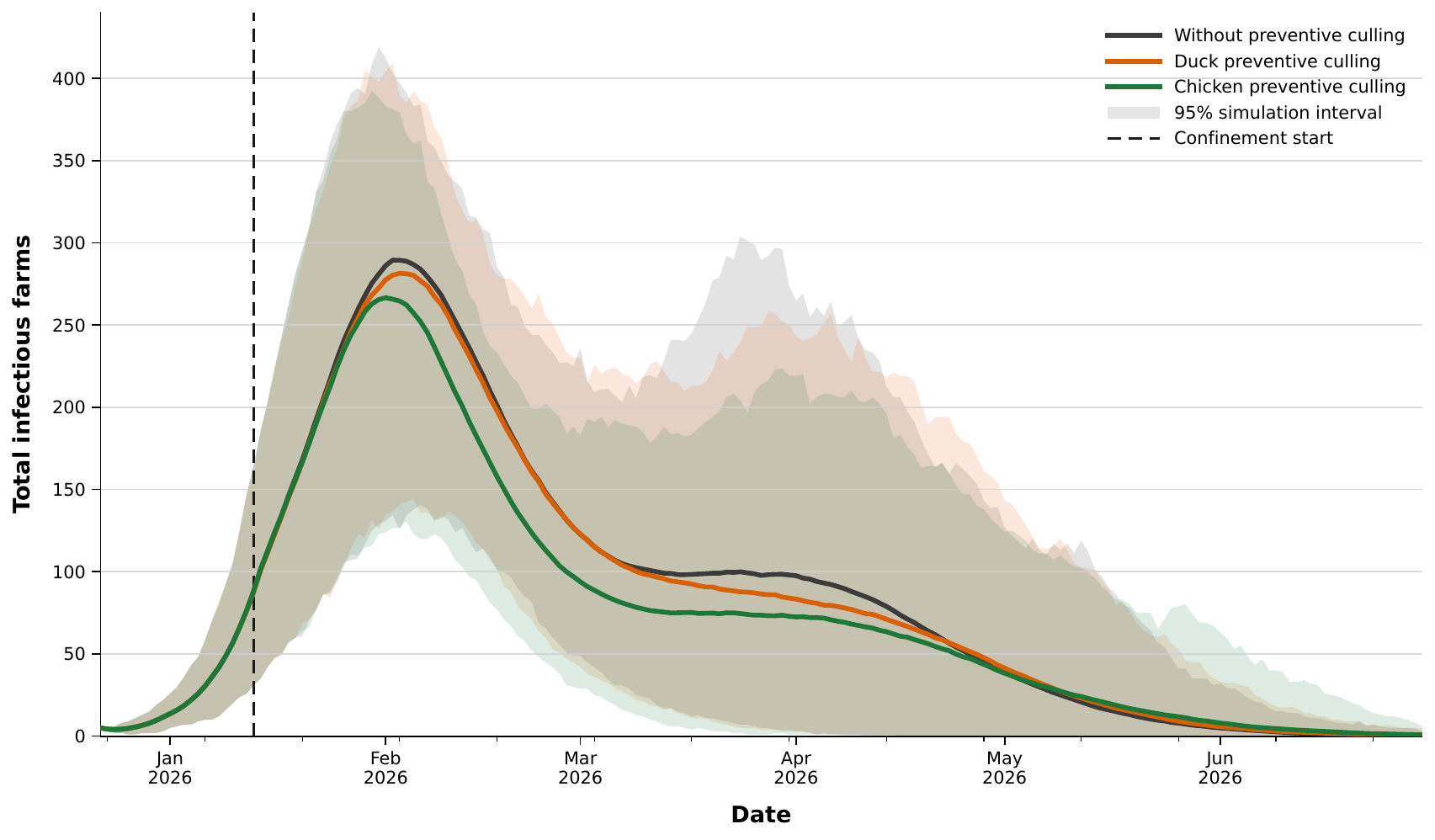} 
\caption{ Baseline confinement of Broiler-2 and organic duck farms. }
\label{fig:targeted_culling_baseline_confinement}
\end{subfigure}
\hfill 
\begin{subfigure}[t]{0.48\textwidth} 
\centering
\includegraphics[ width=\linewidth ]{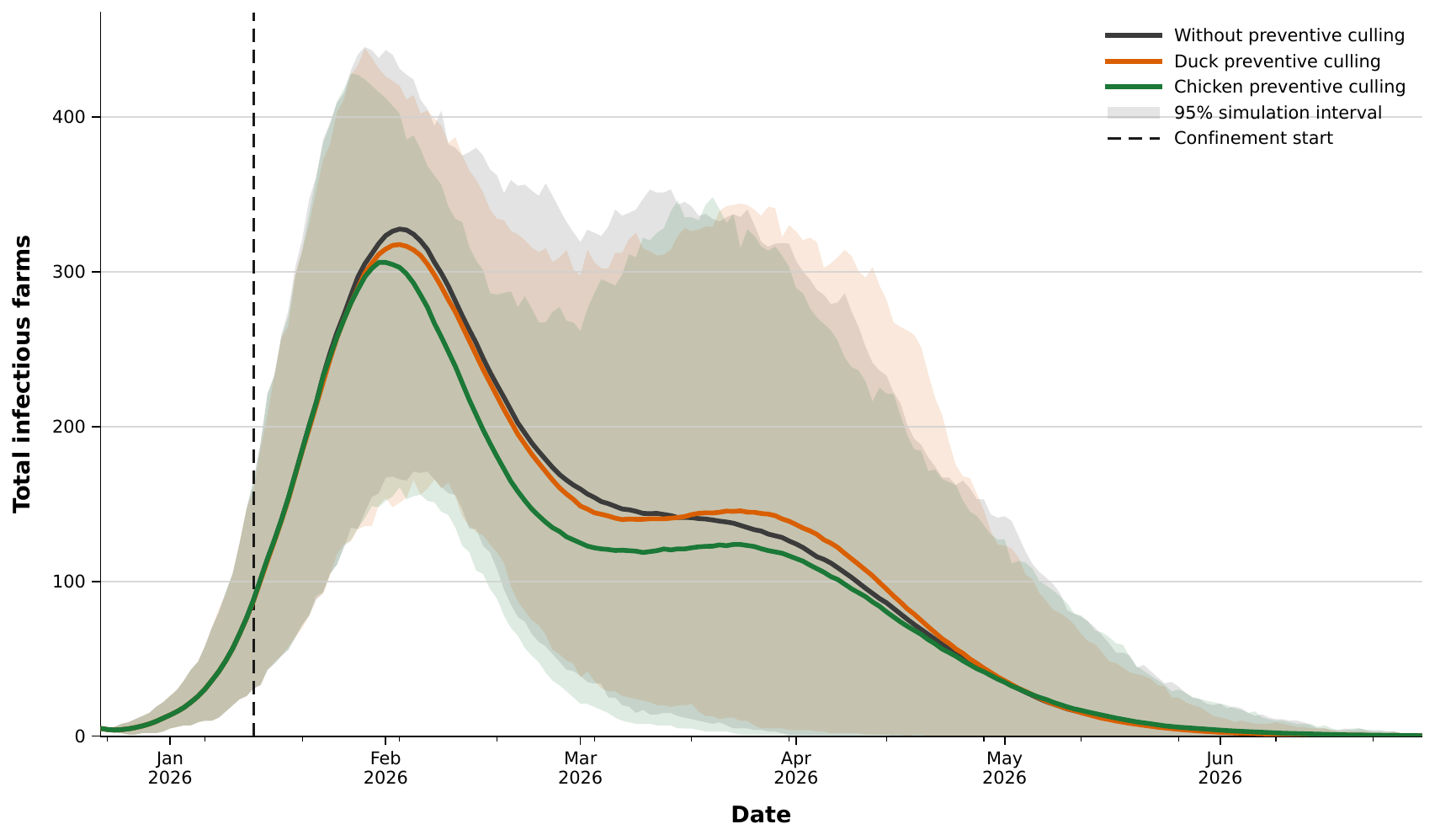} 
\caption{ Broiler-2-only confinement. } 
\label{fig:targeted_culling_broiler_confinement} 
\end{subfigure} 
\vspace{0.5cm} 
\begin{subfigure}[t]{0.48\textwidth} 
\centering \includegraphics[ width=\linewidth ]{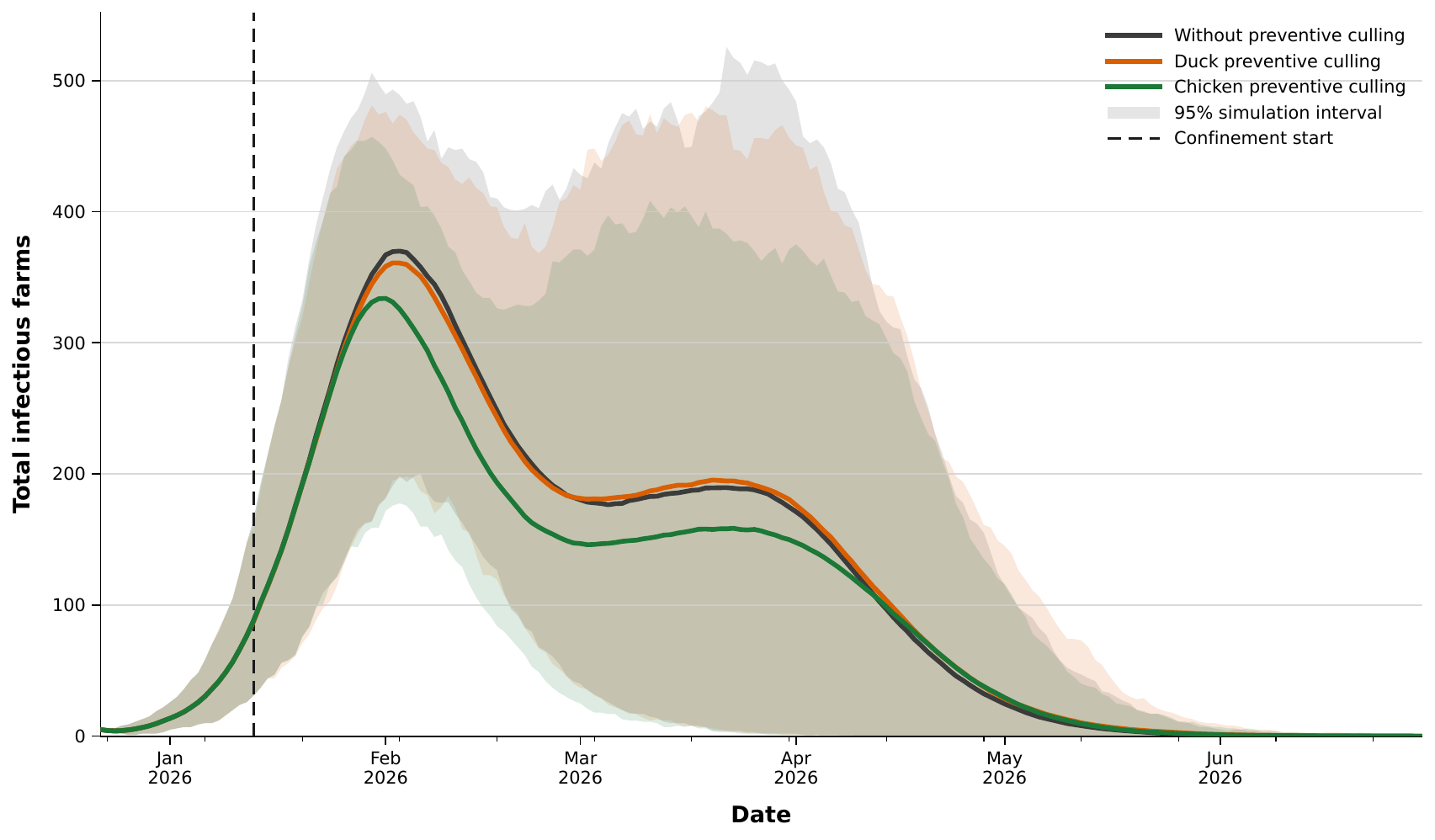}
\caption{ Organic-duck-only confinement. } 
\label{fig:targeted_culling_duck_confinement} 
\end{subfigure} 
\caption{ Simulated HPAI epidemic trajectories under production-targeted preventive-culling policies and alternative confinement configurations. Each panel compares no preventive culling (black), duck-targeted preventive culling (orange), and chicken-targeted preventive culling (green). Panel~(a) shows the baseline configuration in which environmental exposure was interrupted for Broiler-2 and organic duck farms; panel~(b) shows Broiler-2-only confinement; and panel~(c) shows organic-duck-only confinement. Solid lines represent ensemble means, shaded areas represent 95\% simulation intervals, and vertical dashed black lines indicate the confinement start date. }
\label{fig:preventive_culling_comparison} 
\end{figure}

The targeted preventive-culling policies were subsequently evaluated under three confinement configurations: baseline confinement of Broiler-2 and organic duck farms, Broiler-2-only confinement, and organic-duck-only confinement. Figure~\ref{fig:preventive_culling_comparison} compares the resulting temporal epidemic trajectories. 

Across the three confinement configurations, the epidemic followed a broadly similar temporal pattern, with relatively little change in the timing of the principal peak. The main differences were evident in the magnitude of the epidemic and the persistence of infection after the peak. These results suggest that, within the scenarios evaluated, confinement scope affected epidemic magnitude and duration more strongly than epidemic timing. 

Chicken-targeted preventive culling generally produced the lowest ensemble-mean infectious trajectory within each confinement configuration, followed by duck-targeted preventive culling. The no-preventive-culling scenario generally produced the highest and most persistent trajectory. However, the degree of separation among policies varied with confinement scope, indicating that the effect of targeted preventive culling depended on which production classes remained exposed to environmental transmission. 

Under baseline confinement (Figure~\ref{fig:targeted_culling_baseline_confinement}), environmental exposure was interrupted for both Broiler-2 and organic duck farms. Under Broiler-2-only confinement (Figure~\ref{fig:targeted_culling_broiler_confinement}), organic duck and Other farms remained environmentally exposed. Under organic-duck-only confinement (Figure~\ref{fig:targeted_culling_duck_confinement}), Broiler-2 and Other farms remained environmentally exposed. The differences among these panels show that the residual environmental exposure of individual production classes altered the magnitude and persistence of the simulated epidemic. The results are consistent with complementary effects of confinement and preventive culling, confinement reduced environmental exposure for the targeted production classes, whereas preventive culling reduced the number of farms available to participate in subsequent transmission.  

Overall, the targeted-policy analysis indicated a greater reduction in epidemic burden from chicken-targeted preventive culling than from duck-targeted culling alone. This difference was consistent with the larger burden observed among chicken farms and the comparatively small contribution of organic duck farms to the cumulative regional epidemic. The additional effects of confinement timing, environmental-transmission strength, and confinement scope are examined in Section~\ref{sec:confinement_results}.

\subsection{Effects of confinement} \label{sec:confinement_results}

We examined the sensitivity of the simulated epidemic to three aspects of confinement: the timing of confinement, the strength of environmental transmission, and the production classes whose environmental exposure was interrupted. Each analysis varied one component while holding the remaining model assumptions at their baseline values. Table~\ref{tab:sensitivity_summary} summarises the peak ensemble-mean number of infectious farms under the evaluated scenarios. 

\begin{table}[pos=htbp] 
\centering 
\caption{ Sensitivity of the peak ensemble-mean number of infectious farms to confinement timing, environmental-transmission strength, and confinement scope. Unless otherwise stated, confinement began on 14 January 2026, environmental-transmission coefficients were retained at their baseline values, and all recorded preventive culling was included. Environmental transmission multipliers are expressed relative to the baseline values. } 
\label{tab:sensitivity_summary} 
\begin{tabular}{llr} 
\toprule 
\textbf{Analysis} & \textbf{Scenario} & \textbf{Peak mean infectious farms} \\ 
\midrule \multirow{4}{*}{Confinement start date} & 31 December 2025 & 190.30 \\ & 7 January 2026 & 229.25 \\ & 14 January 2026 & 264.52 \\ & 14 February 2026 & 363.73 \\ 
\midrule \multirow{3}{*}{ $\beta_{\mathrm{env},p}$ multiplier } & 0.1 $\times$ baseline & 31.03 \\ & 0.5 $\times$ baseline & 127.52 \\ & 1.0 $\times$ baseline & 264.52 \\
\midrule \multirow{4}{*}{Confinement scope} & Full confinement $(0,0,0)$ & 135.44 \\ & Baseline confinement $(0,0,1)$ & 264.52 \\ & Broiler-2 only $(0,1,1)$ & 296.76 \\ & Organic duck only $(1,0,1)$ & 323.90 \\ 
\bottomrule 
\end{tabular} 
\end{table} 

In the confinement-scope analysis, the vectors in Table~\ref{tab:sensitivity_summary} give the post-confinement environmental--exposure multipliers in the order $(B2,D,O)$. A value of zero indicates complete interruption of the modelled environmental-exposure pathway for the corresponding production class, whereas a value of one indicates that environmental exposure remained unchanged.

\subsubsection{Confinement scope} 

\begin{figure}[pos=htbp]
\centering 
\includegraphics[ width=0.72\textwidth ]{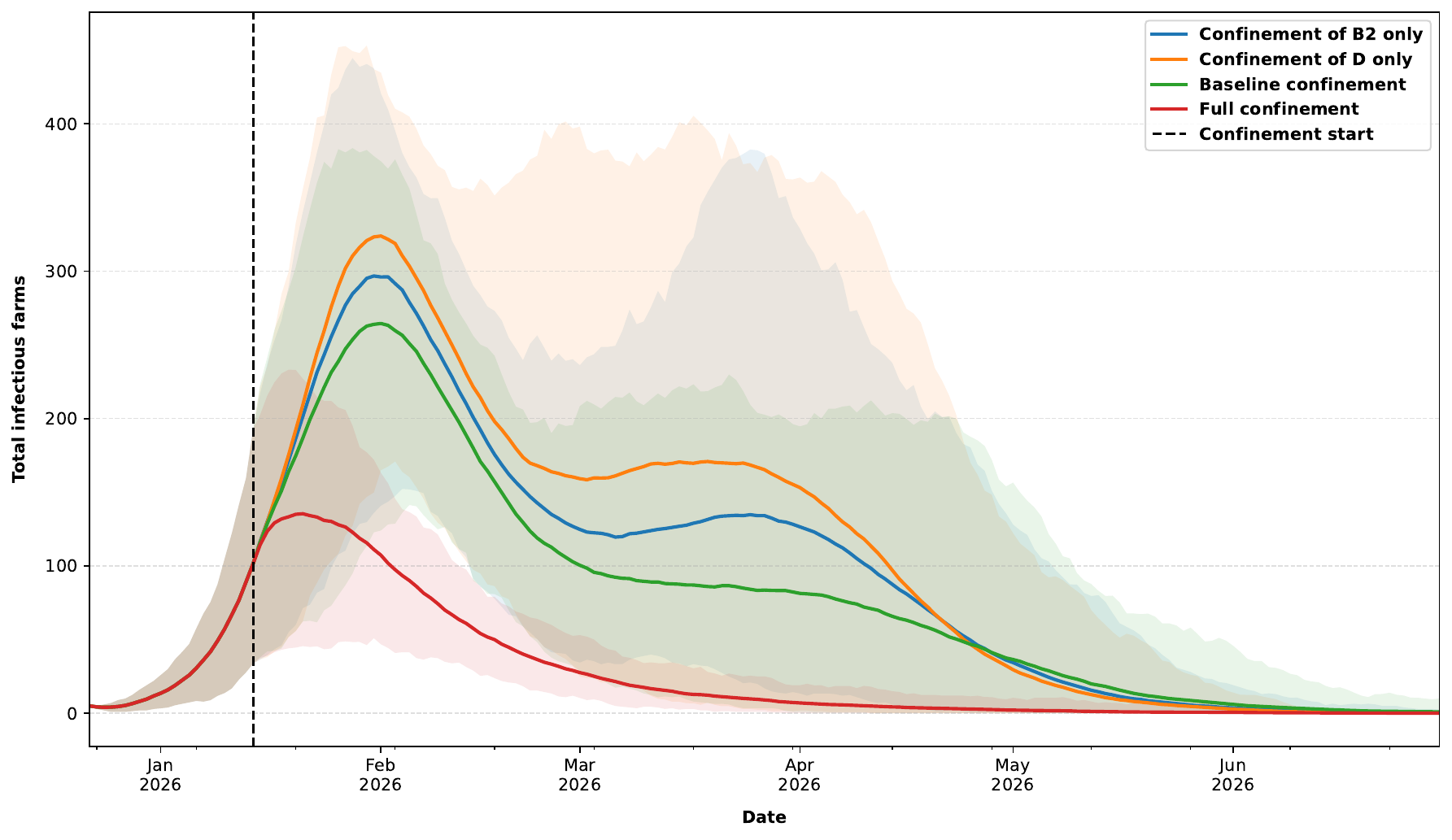} 
\caption{ Simulated epidemic trajectories under alternative production-specific confinement configurations. The scenarios were Broiler-2-only confinement (blue), organic-duck-only confinement (orange), baseline confinement of both Broiler-2 and organic duck farms (green), and full confinement of all three production classes (red). Confinement operated by interrupting the environmental-exposure pathway for the specified production classes. Solid lines represent ensemble means, shaded bands represent 95\% simulation intervals, and the vertical dashed line indicates the confinement start date. Each scenario was based on 100 stochastic simulation runs. } 
\label{fig:single_stratum_confinement}
\end{figure}

The effect of confinement scope is shown in Figure~\ref{fig:single_stratum_confinement}. Restricting environmental exposure for Broiler-2 farms alone produced a lower epidemic peak than restricting exposure for organic duck farms alone. The peak ensemble-mean number of infectious farms was 296.76 under Broiler-2-only confinement, compared with 323.90 under organic-duck-only confinement. The baseline configuration, in which environmental exposure was interrupted for both Broiler-2 and organic duck farms, reduced the peak to 264.52 infectious farms. This represented a reduction of 32.24 infectious farms, or 10.9\%, relative to Broiler-2-only confinement, and a reduction of 59.38 infectious farms, or 18.3\%, relative to organic-duck-only confinement.

Full confinement, in which environmental exposure was interrupted for all three production classes, produced the smallest epidemic peak at 135.44 infectious farms. This was 129.08 fewer infectious farms than under the baseline confinement configuration, corresponding to a 48.8\% reduction in the peak. The reduction under full confinement indicates that continued environmental exposure in the Other production class made a substantial contribution to epidemic magnitude within the model. Among the two single-production confinement strategies, Broiler-2-only confinement was more effective than organic-duck-only confinement. This difference was consistent with the larger epidemic burden observed among Broiler-2 farms relative to organic duck farms. However, neither single-production strategy achieved the reduction observed when both Broiler-2 and organic duck farms were confined, and all three targeted configurations produced larger peaks than full confinement.

\subsubsection{Confinement timing}

\begin{figure}[pos=htbp] 
\centering
\includegraphics[ width=0.72\textwidth ]{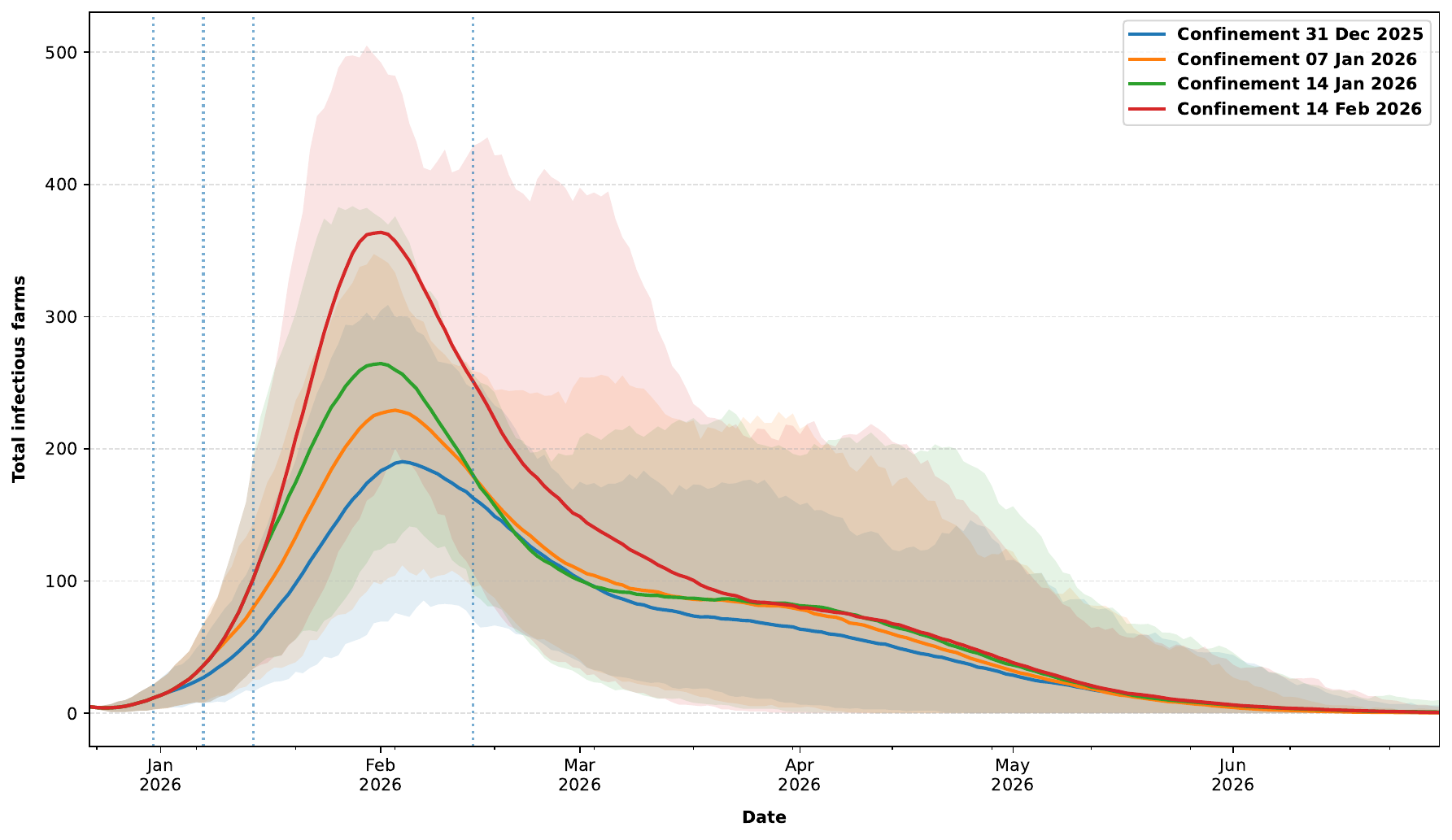}
\caption{ Effect of confinement timing on the simulated epidemic trajectory. Candidate confinement start dates were 31 December 2025 (blue), 7 January 2026 (orange), 14 January 2026 (green), and 14 February 2026 (red). All other model parameters, including environmental-transmission strength and culling policy, were held at their baseline values. Solid lines represent ensemble means, shaded bands represent 95\% simulation intervals, and vertical dotted lines indicate the corresponding confinement start dates. Each scenario was based on 100 stochastic simulation runs. }
\label{fig:timing_confinement} 
\end{figure}

Figure~\ref{fig:timing_confinement} shows that delaying confinement increased the magnitude of the simulated epidemic. When confinement began on 31 December 2025, the peak ensemble-mean number of infectious farms was 190.30. Delaying confinement by one week, to 7 January 2026, increased the peak to 229.25 infectious farms. A further one-week delay, to 14 January 2026, increased the peak to 264.52 infectious farms.

When confinement was delayed until 14 February 2026, the peak increased to 363.73 infectious farms. Relative to confinement beginning on 31 December 2025, this represented an increase of 173.43 infectious farms, or approximately 91.1\%. Thus, delaying confinement by approximately 6.4 weeks nearly doubled the simulated epidemic peak. The trajectories in Figure~\ref{fig:timing_confinement} indicate that earlier confinement reduced epidemic magnitude before the outbreak reached its maximum. Later confinement allowed transmission to continue for longer during the epidemic growth phase, resulting in progressively larger peaks. These results demonstrate that the timing of confinement was an important determinant of epidemic magnitude under the model assumptions.

\subsubsection{Environmental-transmission strength}

\begin{figure}[pos=htbp] 
\centering \includegraphics[ width=0.72\textwidth ]{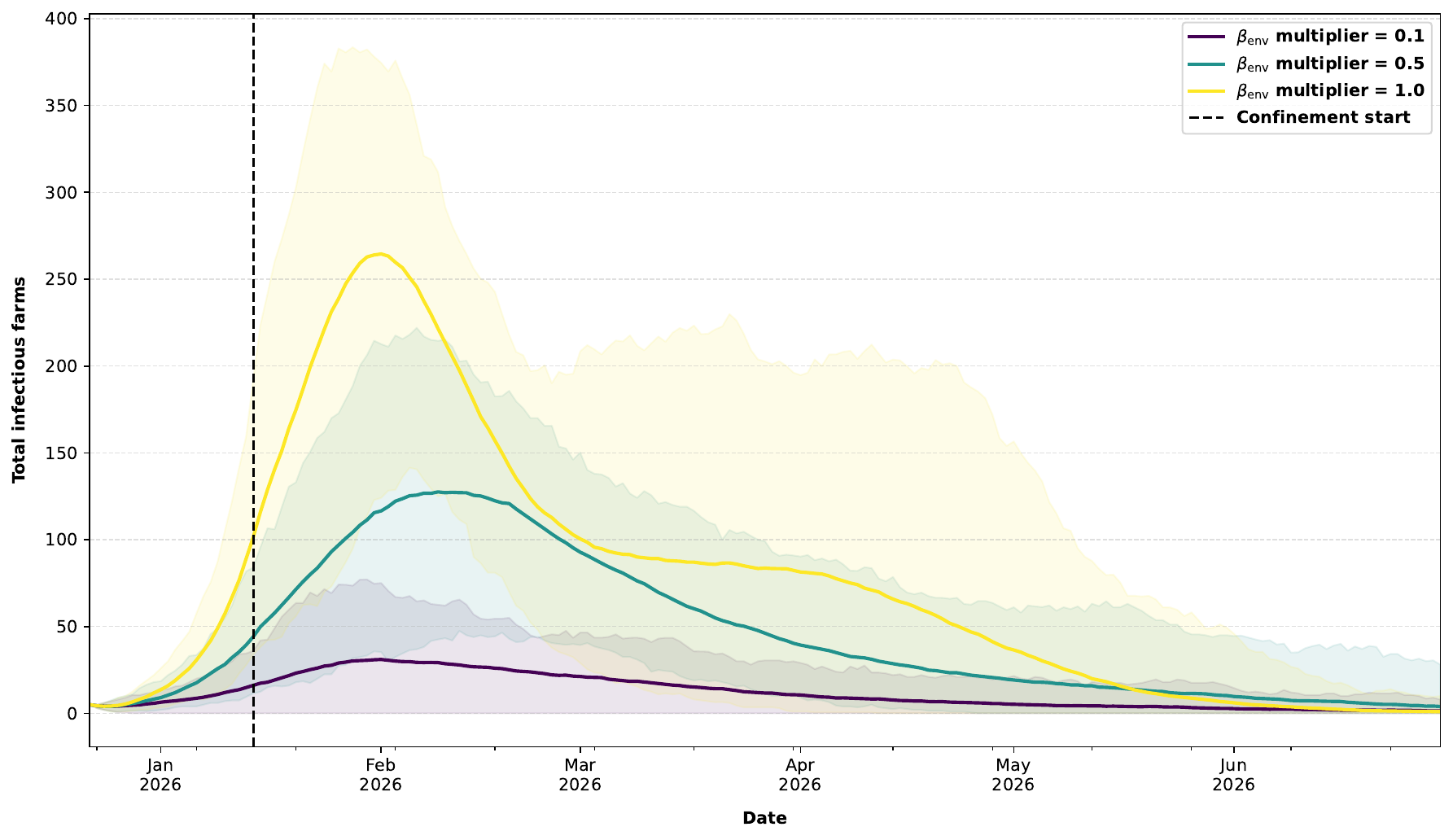}
\caption{ Sensitivity of the simulated epidemic to environmental-transmission strength. The production-specific environmental-transmission coefficients, $\beta_{\mathrm{env},p}$, were multiplied uniformly by 0.1 (purple), 0.5 (yellow), or 1.0 (green), where 1.0 represents the baseline values. Confinement began on 14 January 2026, as indicated by the vertical dashed line. Solid lines represent ensemble means and shaded bands represent 95\% simulation intervals. Each scenario was based on 100 stochastic simulation runs. }
\label{fig:beta_env_sensitivity}
\end{figure}

The simulated epidemic was highly sensitive to the magnitude of the environmental-transmission coefficients (Figure~\ref{fig:beta_env_sensitivity}). When the production-specific coefficients were reduced to 10\% of their baseline values, the peak ensemble-mean number of infectious farms was 31.03. At 50\% of baseline environmental-transmission strength, the peak increased to 127.52 infectious farms. Retaining the full baseline values produced a peak of 264.52 infectious farms. 

Increasing the environmental-transmission multiplier from 0.1 to 0.5 increased the epidemic peak by 96.49 infectious farms. Increasing it from 0.5 to 1.0 produced a further increase of 137.00 infectious farms. The peak under the full baseline environmental-transmission coefficients was more than eight times that obtained when the coefficients were reduced to 10\% of baseline. These results indicate that environmental transmission was a major driver of epidemic amplification within the model. However, the environmental component represented an aggregate pathway and did not distinguish among specific mechanisms such as contaminated equipment, water sources, surfaces, or other environmental contacts. The sensitivity analysis therefore supports the importance of the modelled environmental pathway but does not quantify the contribution of individual real-world environmental transmission routes. 

Overall, the confinement analyses show that epidemic magnitude was sensitive to intervention timing, confinement scope, and the strength of environmental transmission. Earlier confinement produced smaller epidemic peaks, confinement of Broiler-2 farms was more effective than confinement of organic duck farms when only one production class was targeted, and full confinement produced the largest reduction. The results also indicate that the benefit of baseline confinement was limited by continued environmental exposure in the Other production class.

\subsection{Restocking timing and risk of epidemic rebound} \label{sec:restocking}

\begin{figure}[pos=htbp] 
\centering 
\includegraphics[ width=0.75\textwidth ]{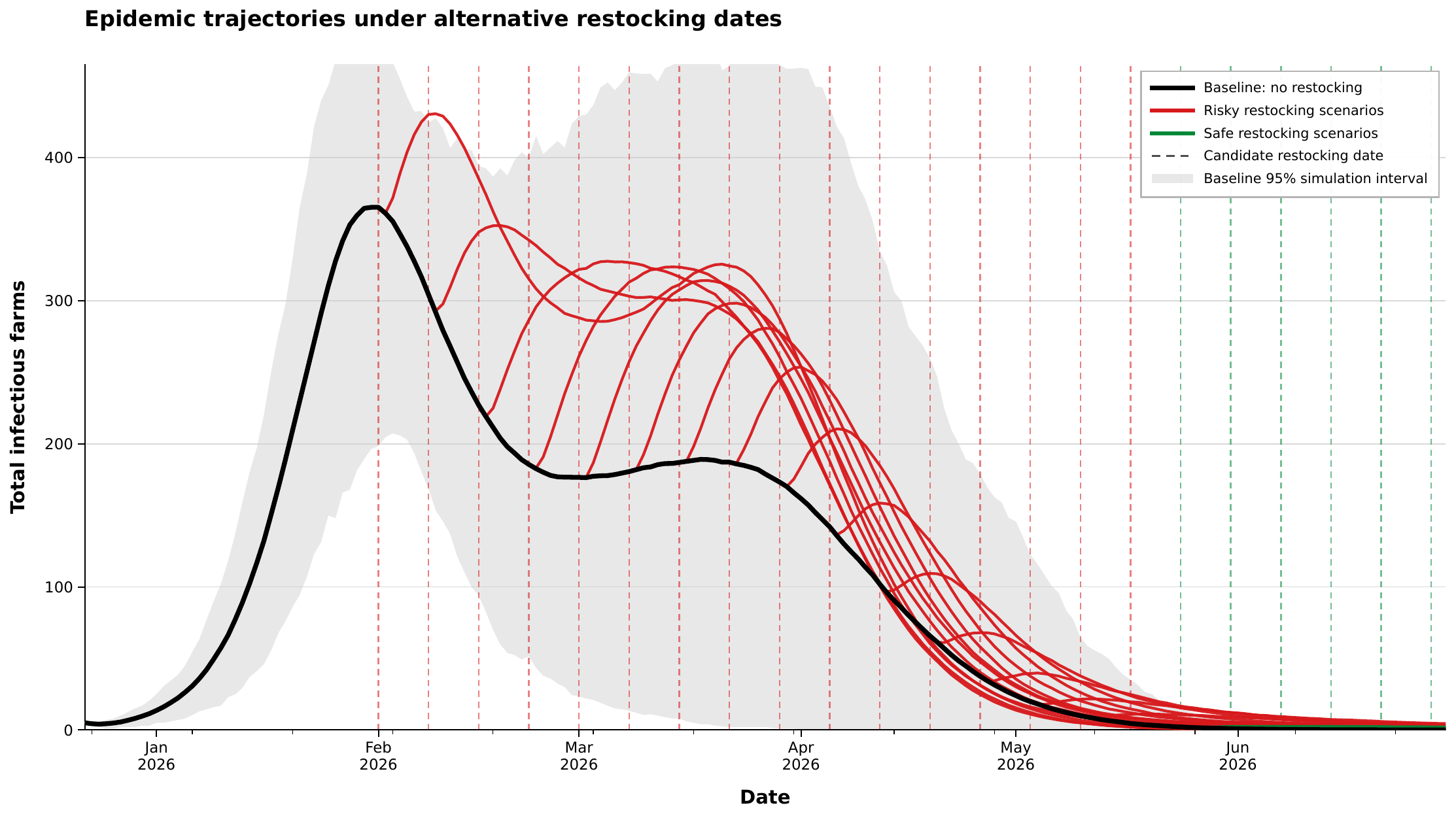} 
\caption{ Simulated epidemic trajectories under alternative candidate restocking dates. The black line represents the ensemble-mean epidemic trajectory without restocking. Red trajectories represent candidate dates classified as risky, whereas green trajectories represent dates satisfying the predefined safety criterion. Vertical dashed lines indicate the candidate restocking dates. The grey shaded area represents the 95\% simulation interval for the no-restocking baseline. }
\label{fig:restocking_trajectories}
\end{figure}

Figure~\ref{fig:restocking_trajectories} shows the simulated epidemic trajectories under the alternative candidate restocking dates. Restocking during the earlier stages of the epidemic produced pronounced secondary increases in the number of infectious farms. As restocking was progressively delayed, the resulting trajectories increasingly converged towards the declining no-restocking baseline, indicating a corresponding reduction in the risk of epidemic rebound. 

\begin{figure}[pos=htbp] 
\centering \includegraphics[ width=0.80\textwidth ]{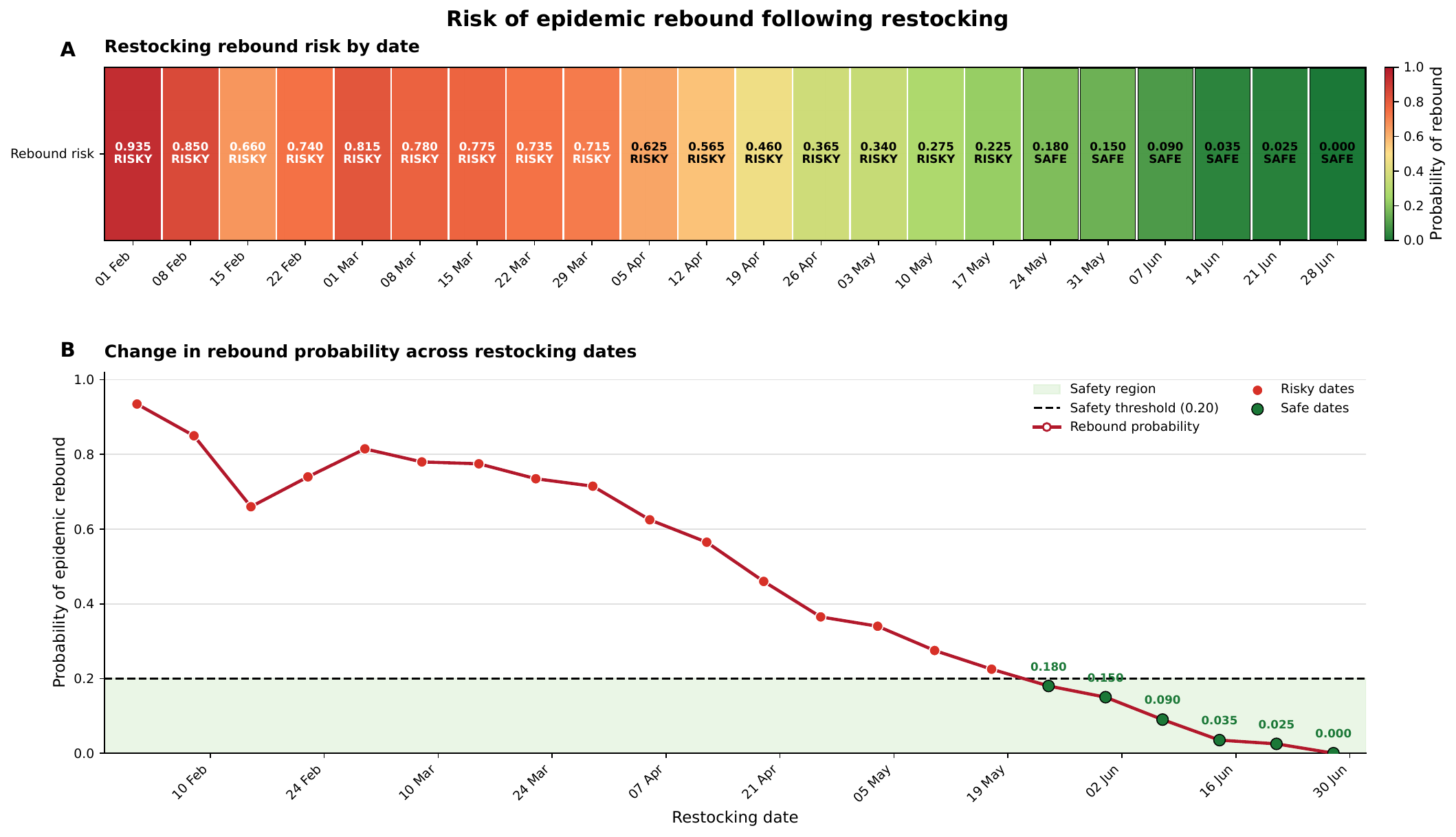} 
\caption{ Estimated probability of epidemic rebound across candidate restocking dates. Panel~(a) presents the estimated rebound probability and corresponding classification for each date. Black borders identify dates satisfying the predefined safety criterion. Panel~(b) shows the temporal change in rebound probability, with the green shaded region indicating probabilities less than or equal to the safety threshold of 0.20. Rebound probability was estimated as the proportion of stochastic runs in which the maximum post-restocking infectious count exceeded the infectious count on the restocking date by more than five farms. } 
\label{fig:restocking_risk}
\end{figure}

The date-specific rebound probabilities are presented in Figure~\ref{fig:restocking_risk}. The estimated rebound probability was 0.935 for restocking on 1 February 2026. Although rebound probability generally declined over time, some short-term variation occurred during February and March. The estimated probabilities were 0.850 on 8 February, 0.660 on 15 February, 0.740 on 22 February, 0.815 on 1 March, 0.780 on 8 March, and 0.775 on 15 March 2026. Rebound probability remained above the predefined safety threshold of 0.20 throughout February, March, and April. It subsequently declined to 0.340 on 3 May, 0.275 on 10 May, and 0.225 on 17 May 2026. The first candidate restocking date satisfying the safety criterion was 24 May 2026, when the estimated rebound probability decreased to 0.180 (Figure~\ref{fig:restocking_risk}). 

The estimated rebound probability decreased further to 0.150 on 31 May, 0.090 on 7 June, 0.035 on 14 June, 0.025 on 21 June, and 0.000 on 28 June 2026. Candidate restocking dates from 1 February to 17 May were therefore classified as risky, whereas all candidate dates from 24 May to 28 June were classified as safe. Taken together, Figures~\ref{fig:restocking_trajectories} and~\ref{fig:restocking_risk} show that the simulated probability of resurgence decreased substantially as restocking was delayed. Restocking before late May reintroduced susceptible farms while sufficient infection remained to generate renewed epidemic growth. From 24 May onward, the estimated rebound probability remained below the predefined threshold and continued to decline as the residual infectious burden approached zero. Because observed outbreak and intervention records were available only through 7 April 2026, the restocking classifications for May and June represent model-based projections rather than observed post-restocking outcomes. 

\subsubsection{Sensitivity to restocking intensity} \label{sec:restocking_fraction_results} 

\begin{figure}[pos=htbp] 
\centering \includegraphics[ width=0.72\textwidth ]{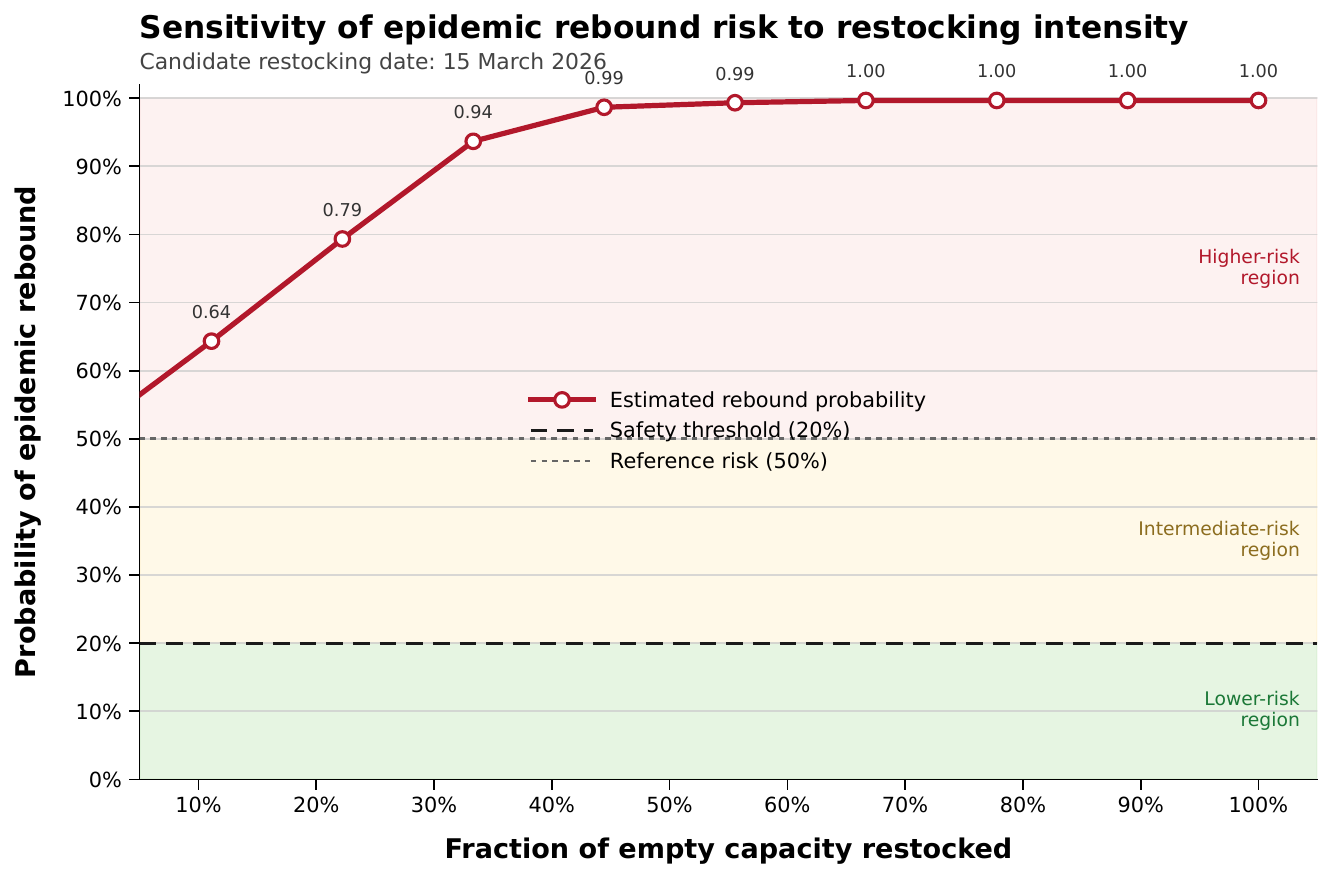} 
\caption{ Sensitivity of epidemic rebound probability to the proportion of available capacity restored on 15 March 2026. Restocking fractions ranged from 10\% to 100\% of available capacity. The dashed black line indicates the predefined safety threshold of 20\%, and the dotted grey line indicates a 50\% reference probability. None of the evaluated restocking fractions satisfied the safety criterion on this candidate restocking date. } 
\label{fig:restocking_fraction} 
\end{figure}

The relationship between restocking intensity and rebound probability is shown in Figure~\ref{fig:restocking_fraction}. Restocking intensity strongly affected rebound risk when restocking was implemented during the active epidemic period. For the candidate restocking date of 15 March 2026, restoring 10\% of available capacity resulted in an estimated rebound probability of approximately 0.64. The estimated probability increased to approximately 0.79 when 20\% of available capacity was restored and to 0.94 at a restocking fraction of 30\%. 

The rebound probability reached approximately 0.99 at restocking fractions of 40\% and 50\%, and approached 1.00 when 60\% or more of available capacity was restored. None of the restocking fractions evaluated on 15 March satisfied the predefined safety criterion of 0.20 (Figure~\ref{fig:restocking_fraction}). The sensitivity analysis showed that rebound probability was determined by both the timing and intensity of restocking. Although smaller restocking fractions produced lower rebound probabilities than more extensive repopulation, reducing the fraction to 10\% was insufficient to make restocking on 15 March epidemiologically safe under the predefined criterion. The rebound probability at this fraction remained more than three times the safety threshold. 

\subsubsection{Comparison of original and capacity-based restocking} \label{sec:restocking_formulation_results} 

\begin{figure}[pos=htbp] 
\centering \includegraphics[ width=0.78\textwidth ]{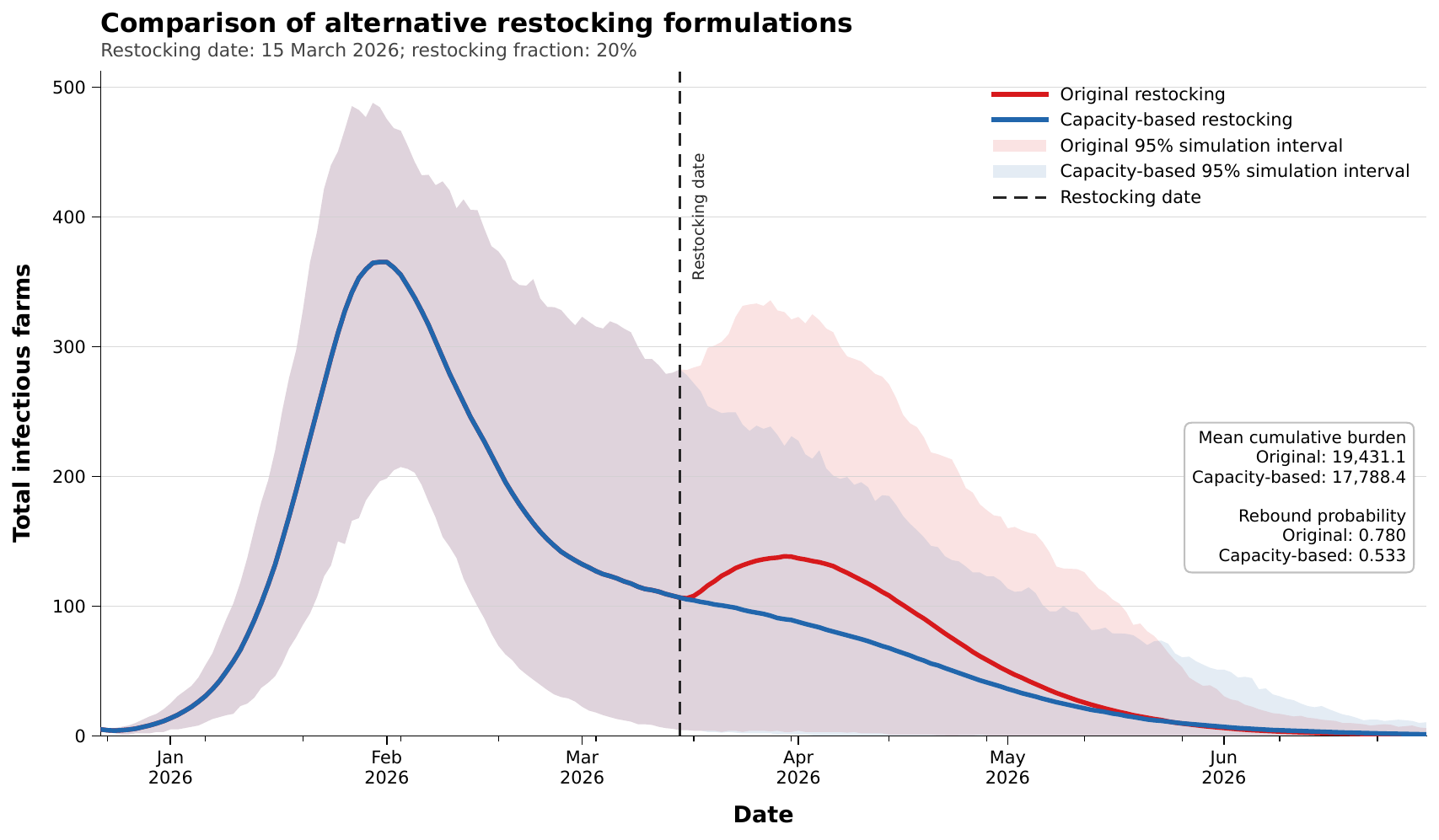} 
\caption{ Comparison of epidemic trajectories under the original and capacity-based restocking formulations. Restocking was implemented on 15 March 2026 at a fraction of 20\%. Solid lines represent ensemble-mean infectious trajectories, and shaded areas represent 95\% simulation intervals. The vertical dashed line indicates the restocking date. Capacity-based restocking reduced the mean cumulative burden from 19,431.1 to 17,788.4 infectious-farm-days and reduced the estimated rebound probability from 0.780 to 0.533. }
\label{fig:restocking_formulations}
\end{figure}

Figure~\ref{fig:restocking_formulations} compares the epidemic trajectories generated by the original and capacity-based restocking formulations. Both formulations were evaluated using a restocking date of 15 March 2026 and a restocking fraction of 20\%. The original formulation produced a mean cumulative burden of 19,431.1 infectious-farm-days, whereas the capacity-based formulation produced a mean cumulative burden of 17,788.4 infectious-farm-days. The absolute difference was 1,642.7 infectious-farm-days, corresponding to an 8.45\% reduction relative to the original formulation. 

The estimated rebound probability decreased from 0.780 under the original formulation to 0.533 under capacity-based restocking. This represented an absolute reduction of 0.247, equivalent to 24.7 percentage points, and a relative reduction of approximately 31.7\%. Despite this reduction, the rebound probability under capacity-based restocking remained above the predefined safety threshold of 0.20. Restricting restocking to available capacity therefore reduced, but did not eliminate, the rebound risk associated with restocking on 15 March. 

The higher burden under the original formulation resulted from applying the restocking fraction directly to the baseline county--production population, regardless of the number of farms still present at the restocking date. In contrast, the capacity-based formulation applied the fraction only to the difference between the current and baseline populations. Consequently, the capacity-based formulation prevented county--production populations from exceeding their baseline capacities and generated lower post-restocking epidemic trajectories. 

Overall, the alternative restocking analyses yielded three principal findings. First, rebound probability declined as restocking was delayed, with 24 May 2026 representing the first candidate date satisfying the predefined safety criterion. Second, reducing the restocking fraction alone did not make the early candidate date of 15 March safe. Third, the capacity-based formulation produced lower cumulative burden and rebound probability than the original formulation, although the risk on 15 March remained above the safety threshold.

\section{Discussion} \label{sec:discussion}

We developed a stochastic, spatially explicit, county--production metapopulation model to evaluate HPAI control and poultry restocking on Jolly Island. The model incorporated local, environmental, movement-mediated, and spatial transmission together with reactive and preventive culling, production-specific confinement, and capacity-based restocking. The results demonstrated substantial geographic and production-specific heterogeneity, reductions in epidemic burden under preventive culling and early confinement, and a strong dependence of post-restocking rebound risk on the timing and intensity of repopulation.

\subsection{Epidemic heterogeneity and control effectiveness} 

The simulated epidemic burden was concentrated primarily in Berks, Indiana, and Lancaster, while most counties experienced substantially lower burdens. The spatiotemporal maps showed that transmission intensified within a broadly contiguous geographic cluster before declining, without producing similarly large new foci elsewhere. This concentration was consistent with the structured transmission processes represented in the model, although the relative contributions of local, environmental, movement-mediated, and spatial transmission could not be separated from the maps alone.

Burden also differed substantially among production classes. Other production systems generated the greatest cumulative burden and accounted for much of the persistent transmission during March and April. Broiler-2 farms also contributed substantially, whereas organic duck farms experienced a smaller and earlier-resolving epidemic. These differences reflected the geographic distribution and number of farms, recorded movement patterns, environmental shedding, and production-specific transmission parameters, rather than biological susceptibility alone. Nevertheless, they demonstrate the importance of explicitly representing production structure when evaluating regional control strategies.

Including all recorded preventive culling reduced the overall cumulative burden by 16.7\%, equivalent to approximately 2,731 infectious-farm-days averted. The proportional reduction was greatest among Broiler-2 farms, while Other production systems accounted for the largest absolute reduction. The effect among organic duck farms was comparatively small. Chicken-targeted preventive culling also produced a greater reduction than duck-targeted culling, suggesting that directing interventions towards production systems contributing substantially to epidemic burden may provide greater regional benefit.

These epidemiological benefits should, however, be considered alongside the broader consequences of preventive depopulation. The analysis did not quantify the number of uninfected farms culled, birds destroyed, compensation costs, production losses, animal-welfare consequences, or effects on food supply. A strategy that reduces infectious-farm-days may not remain optimal when these additional outcomes are included.

The confinement analyses similarly demonstrated the importance of both intervention timing and scope. Delaying confinement from 31 December 2025 to 14 February 2026 increased the simulated epidemic peak by approximately 91\%. Within the model, earlier confinement interrupted the designated environmental-exposure pathways while the epidemic was still in its growth phase, whereas later implementation permitted transmission to continue for longer before the intervention became active.

When only one production class was confined, Broiler-2 confinement produced a smaller epidemic peak than organic-duck confinement. Confining both Broiler-2 and organic duck farms produced a greater reduction than either single-production strategy, while confinement of all production classes produced the smallest peak. The additional benefit of full confinement indicated that continued environmental exposure among Other farms contributed substantially to epidemic amplification.

The epidemic was also highly sensitive to the environmental-transmission coefficients. Reducing these coefficients to 10\% of their baseline values produced a much smaller epidemic than retaining the complete baseline values. This finding supports the importance of reducing environmental exposure but does not identify which specific measure would be most effective. The environmental compartment represented an aggregate pathway and did not distinguish among contaminated water, litter, equipment, vehicles, surfaces, or wild-bird interfaces.

Overall, these results indicate that HPAI control should account for both geographic and production-specific heterogeneity. Rapid confinement and targeted preventive culling may provide greater benefit when directed towards high-burden locations and production systems. However, the operational value of these measures will also depend on feasibility, adherence, surveillance capacity, economic costs, and animal-welfare considerations.

\subsection{Restocking and post-epidemic recovery} The restocking analysis demonstrated that poultry reintroduction should be treated as an epidemiological intervention rather than solely as an economic recovery decision. Restocking during the active epidemic replenished the susceptible farm population while sufficient infection remained to generate secondary epidemic growth. As restocking was delayed, the resulting trajectories increasingly converged towards the declining no-restocking baseline. 

Under the model assumptions, 24 May 2026 was the first candidate date satisfying the predefined rebound-probability threshold of 0.20, with an estimated probability of 0.180. The probability continued to decline during June and reached zero for the final candidate date of 28 June 2026. This model-estimated date should not be interpreted as a universal calendar rule. It was conditional on the model structure, parameter values, intervention schedules, restocking fraction, rebound definition, and selected decision threshold.

Restocking intensity also influenced rebound risk. When restocking was fixed at 15 March 2026, none of the tested fractions met the safety criterion, including restoration of only 10\% of available capacity. Increasing the fraction further increased the probability of resurgence, which approached one when 60\% or more of available capacity was restored. Reducing the restocking fraction therefore could not compensate fully for repopulation during the active epidemic period. Capacity-based restocking reduced cumulative burden by 8.45\% and decreased rebound probability from 0.780 to 0.533 relative to the original formulation. It also prevented county--production populations from exceeding their baseline capacities. Nevertheless, the rebound probability remained above the safety threshold on 15 March, demonstrating that correcting the restocking formulation reduced risk but did not make this early date safe. 

These findings support a phased, risk-based approach to recovery. Initial restocking could be restricted to a proportion of verified empty capacity after the residual infection level has declined sufficiently, with later expansion conditional on continued surveillance. Operational decisions should also incorporate cleaning and disinfection, environmental testing, time since the most recent outbreak, farm-level biosecurity, movement restrictions, and regulatory requirements. 

The rebound threshold of 0.20 provided a transparent decision rule but was not derived from an economic optimisation or formal utility analysis. A lower threshold would delay restocking and reduce the accepted probability of resurgence, whereas a higher threshold would permit earlier recovery at the cost of greater epidemiological risk. The appropriate threshold should therefore reflect decision-makers' tolerance for epidemiological, economic, and operational consequences.

\subsection{Limitations} 

The county--production structure provided a practical representation of geographic and production-specific heterogeneity, but it required farms within each stratum to be treated as epidemiologically homogeneous. Farm-level variation in flock size, housing, biosecurity, management, susceptibility, and contact behaviour was therefore not explicitly represented. Environmental contamination was aggregated at county level, and the model did not distinguish among individual environmental transmission routes. Recorded farm movements also excluded unobserved contacts involving personnel, vehicles, equipment, and informal movements. Some of these contacts may have been represented indirectly through the local, environmental, or spatial components, but their individual contributions could not be estimated. 

Culling and restocking were simplified to accommodate the aggregated model structure. The epidemiological state of each scheduled culled farm was not available, so recorded culls were allocated sequentially across the susceptible, exposed, infectious, and removed compartments. Restocked farms were assumed to be susceptible and uninfected, without explicit representation of cleaning, disinfection, environmental testing, partial flock replacement, or enhanced post-restocking biosecurity. The estimated rebound probabilities should therefore be interpreted as conditional scenario outcomes rather than direct operational recommendations. 

The model parameters were specified as scenario values rather than formally estimated from the observed epidemic. Consequently, the 95\% simulation intervals represent stochastic variation under fixed parameters and do not capture the full uncertainty associated with parameter selection, model structure, or incomplete observation of the transmission network. Although sensitivity analyses were conducted for environmental transmission, confinement timing, confinement scope, and restocking intensity, comprehensive uncertainty quantification would require formal calibration and a broader global sensitivity analysis. 

The epidemic was seeded using the first five confirmed farms without allowing subsequent external introductions. Repeated introductions from wild birds or other sources could prolong transmission and delay the estimated restocking window. Moreover, observed outbreak and intervention records were available only through 7 April 2026, whereas the simulation continued until 30 June 2026. The projected decline after 7 April and the estimated first safe candidate date of 24 May were therefore model-based projections conditional on the represented transmission and control processes remaining applicable beyond the observation period.

Despite these limitations, the model provides an internally consistent framework for comparing HPAI control and recovery strategies under spatial and stochastic uncertainty. Its principal value lies in evaluating the relative consequences of alternative policies rather than predicting the infection status of individual farms or providing an unconditional calendar date for operational restocking.

\section{Conclusion} \label{sec:conclusion}

This study developed a stochastic, spatially explicit, county--production metapopulation model for evaluating HPAI control and poultry restocking on Jolly Island. The model showed that epidemic burden was geographically concentrated and differed substantially among poultry production systems. Preventive culling reduced cumulative infectious-farm-days, while earlier and broader confinement reduced epidemic magnitude. Environmental transmission was also an important driver of epidemic amplification within the model. 

Restocking outcomes depended on both the timing and intensity of repopulation. Under the model assumptions, 24 May 2026 was the first candidate date satisfying the predefined rebound-probability threshold. Capacity-based restocking produced a lower cumulative burden and rebound probability than applying the restocking fraction directly to the baseline farm population, although it did not make early restocking safe. 

By integrating active epidemic control with post-outbreak recovery, the framework provides a consistent approach for comparing intervention and restocking strategies under spatial and stochastic uncertainty. The findings support timely epidemic control, geographically and production-targeted interventions, and phased restocking based on residual epidemiological risk. However, model outputs should be considered alongside surveillance, cleaning and disinfection, farm biosecurity, economic and animal-welfare considerations, and regulatory requirements when informing operational decisions.

\section*{Funding}
The article processing charge (APC) is supported by H.O.F and R.H.L.I's Journal Publication Fund provided by the School of Engineering, Computer and Mathematical Sciences, Auckland University of Technology.

\section*{Declaration of competing interest}
The authors declare that they have no known competing financial interests or personal relationships that could have appeared to influence the work reported in this paper. 

\section*{Data availability}
The datasets analysed during the current study and codes are
available in the GitHub repository--\href{https://github.com/hamfat/HPAI-Challenge}{Flockbusters}

\section*{Acknowledgements}
We thank the organisers of the HPAI Modelling Challenge for developing this interesting modelling problem and for the opportunity to participate in the challenge. 
%\appendix

\printcredits

%\bibliographystyle{cas-model2-names}

%\bibliography{cas-refs}
%\bibliography{references}

\begin{comment}
\bio{}
Author biography without author photo.
Author biography. Author biography. Author biography.
Author biography. Author biography. Author biography.
Author biography. Author biography. Author biography.
Author biography. Author biography. Author biography.
Author biography. Author biography. Author biography.
Author biography. Author biography. Author biography.
Author biography. Author biography. Author biography.
Author biography. Author biography. Author biography.
Author biography. Author biography. Author biography.
\endbio

\bio{figs/pic1}
Author biography with author photo.
Author biography. Author biography. Author biography.
Author biography. Author biography. Author biography.
Author biography. Author biography. Author biography.
Author biography. Author biography. Author biography.
Author biography. Author biography. Author biography.
Author biography. Author biography. Author biography.
Author biography. Author biography. Author biography.
Author biography. Author biography. Author biography.
Author biography. Author biography. Author biography.
\endbio

\bio{figs/pic1}
Author biography with author photo.
Author biography. Author biography. Author biography.
Author biography. Author biography. Author biography.
Author biography. Author biography. Author biography.
Author biography. Author biography. Author biography.
\endbio
\end{comment}

\end{document}